\documentclass{iopjournal}
\usepackage{amsmath,amssymb,amsthm,mathtools}
\usepackage{float}
\usepackage{booktabs}
\usepackage{placeins}
\usepackage{caption}
\usepackage{url}
\usepackage[numbers,sort&compress]{natbib}
\makeatletter

\fancyhfoffset[L]{0mm}
\fancypagestyle{plain}{\fancyhf{}\renewcommand{\headrulewidth}{0pt}\renewcommand{\footrulewidth}{0pt}}
\renewcommand{\headrulewidth}{0pt}
\renewcommand{\footrulewidth}{0pt}

\renewcommand{\articletype}[1]{}

\renewcommand{\marginpar}[1]{}
\renewcommand\section{\@startsection {section}{1}{\z@}%
                   {-4ex \@plus -1.5ex \@minus -.3ex}%
                   {2ex \@plus .3ex}%
                   {\reset@font\large\bfseries\raggedright}}
\renewcommand\subsection{\@startsection{subsection}{2}{\z@}%
                   {-3.5ex \@plus -1ex \@minus -.2ex}%
                   {1.5ex \@plus .2ex}%
                   {\reset@font\normalsize\bfseries\raggedright}}
                   
\makeatother

\newtheorem{theorem}{Theorem}[section]
\newtheorem{proposition}{Proposition}[section]
\newtheorem{remark}{Remark}[section]

\begin{document}
\thispagestyle{plain}

\title{Absence of critical scaling in the Schelling segregation model}

\author{Mouhssine Rifaki$^{1,*}$}

\affil{$^1$Stanford University, Stanford, CA 94305, USA}
\affil{$^*$Corresponding author.}

\email{rifaki@stanford.edu}

\keywords{Schelling segregation model, finite-size scaling, sub-critical cascades, agent-based models}

\begin{abstract}
\noindent
We find no evidence of critical scaling in the Schelling segregation model, in either the Moore neighborhood or its dense-spectrum extension to Chebyshev radii up to $r_0 = 6$ ($k = 168$ neighbors). On periodic grids up to $L = 320$ with 50 trials per point ($> 12\,500$ runs in total), every finite-size scaling diagnostic in the Moore baseline fails: the per-$L$ $T_c$ does not drift, $\mathrm{Var}(S) \sim L^{-2.02 \pm 0.09}$ matches trivial averaging, $\gamma/\nu \approx 0$, and the scaling collapse never reaches a finite optimum. The 8-site Moore neighborhood restricts satisfaction to ratios $j/k$ with $k \leq 8$, giving $S(T)$ a staircase structure with 23 rational thresholds; discreteness alone does not forbid criticality (cf.\ the Ising model), but the scaling evidence rules it out empirically for this model. A branching-ratio calculation predicts subcritical cascades of mean size $1/(1-R)$ and is validated by perturbation experiments to within 15\%; the multiscalar dissimilarity length stays finite across the transition. The dense-spectrum extension strengthens the negative verdict rather than approaching criticality: across $r_0 \in \{3,4,5,6\}$ on $L \in \{40, 80, 160\}$ the Binder cumulant has no $L$-curve crossing and the per-$L$ $T_c$ drift is monotonic and unsaturated; at $r_0 = 4$, extending to $L = 320$ gives a 4-point variance exponent $\alpha = -2.70$, well below the critical boundary $\alpha = -2$, dissolving an apparent super-critical signal of $\alpha = +0.81$ visible only on $L \in \{40, 80\}$; the peak $|dS/dT|$ at $L = 40$ decreases monotonically from $97.5$ ($r_0 = 1$) to $39.1$ ($r_0 = 6$), the opposite of the sharpening expected at criticality. The mechanism is the absence of long-range correlation in equilibrium plus deterministic high-$k$ dynamics (seed-to-seed move-count standard deviation collapses from $1\,557$ at $r_0 = 4$ to $\sim 100$ at $r_0 \geq 5$), and not the staircase structure that pins the Moore baseline. With a Beta-distributed heterogeneous tolerance, the intolerant tail drives segregation even at moderate population-average tolerance. The staircase theorem and cascade mechanism together account for the Schelling transition without invoking critical phenomena.
\end{abstract}

\section{Introduction}

In Schelling's model \citep{schelling1971,schelling1978}, agents of two types on a lattice move whenever too few of their neighbors are like them. Despite the simplicity of the rule, the model produces large-scale spatial sorting and has been studied extensively in physics \citep{dall2008,gauvin2009,castellano2009}, economics \citep{pancs2007,grauwin2012}, and geography \citep{clark2008,hatna2012}.

Several authors have asked whether the mixed-to-segregated transition belongs to a universality class. \citet{gauvin2009} ran FSS on grids up to $L = 60$ and reported exponents close to 2D Ising. \citet{stauffer2007} drew similar analogies. \citet{dall2008} pointed out that the dynamics breaks detailed balance, and \citet{vinkovicdostanic2006} proposed a surface-tension interpretation.

This paper revisits the question at larger scale ($L$ up to 320, 50 trials per point, over $12\,500$ runs distributed across CI workers). The Moore neighborhood has 8 sites, so satisfaction is always a ratio $j/k$ with $k \leq 8$, and the order parameter inherits a staircase structure pinned to 23 rational thresholds. Discrete local states do not by themselves forbid criticality (the Ising model is the obvious counterexample), but here every scaling diagnostic (variance exponents, susceptibility, Binder crossings, data collapse) is inconsistent with a continuous transition. We complement the FSS analysis with a cascade branching-ratio calculation validated by perturbation experiments, the multiscalar dissimilarity framework of \citet{randonfurling2020}, a multi-radius extension probing Chebyshev neighborhoods up to $k = 168$, and a heterogeneous-tolerance variant using Beta-distributed thresholds.

\section{Model and Observables}
\label{sec:model}

\subsection{Schelling Dynamics}

Consider an $L \times L$ square lattice with periodic boundary conditions. Each site is empty or occupied by an agent of type $A$ or $B$. The occupied fraction is $\rho$ and the type-$A$ fraction among occupied sites is $f_A$. We fix $\rho = 0.9$ and $f_A = 0.5$ throughout, following the standard benchmark in the Schelling literature \citep{dall2008,gauvin2009}. The high density ensures that vacancy-mediated dynamics dominates over free diffusion, while equal type fractions eliminate trivial majority/minority asymmetries.

Each agent $i$ at site $(r,c)$ computes its local satisfaction
\begin{equation}
s_i \;=\; \frac{\text{number of same-type occupied Moore neighbors}}{\text{number of occupied Moore neighbors}}\,,
\label{eq:satisfaction}
\end{equation}
and is satisfied when $s_i \geq T$ for a global tolerance threshold $T \in [0,1]$. At each discrete time step, all unsatisfied agents are identified, shuffled into a random order, and relocated sequentially to uniformly chosen empty sites. The grid is updated after each individual move, so that later agents within the same step see the effects of earlier relocations. This asynchronous update matches Schelling's original specification.

The dynamics halt when no agent moves, or when the segregation index (defined below) stabilizes to within a standard deviation of $10^{-3}$ over 20 consecutive steps, or after 2000 steps.

\subsection{Order Parameters}

The primary order parameter is the normalized segregation index
\begin{equation}
S \;=\; \frac{\bar{s} - s_{\mathrm{rand}}}{1 - s_{\mathrm{rand}}}\,,
\label{eq:segindex}
\end{equation}
where $\bar{s}$ is the satisfaction averaged over all occupied sites and $s_{\mathrm{rand}} = f_A^2 + f_B^2$ is the expected satisfaction under a uniformly random assignment of types to occupied sites. By construction, $S = 0$ for a random configuration and $S \to 1$ for full segregation.

We also track the interface density
\begin{equation}
I \;=\; \frac{\text{number of unlike nearest-neighbor occupied pairs}}{\text{total number of occupied nearest-neighbor pairs}}\,,
\label{eq:interface}
\end{equation}
counted over horizontal and vertical bonds. This quantity measures local mixing at cluster boundaries: $I \to 0$ in the segregated phase and $I \to 2f_A f_B = 0.5$ for a well-mixed configuration at $f_A = 0.5$.

For the FSS analysis, we use the susceptibility $\chi = L^2\,\mathrm{Var}(S)$ and the Binder cumulant $U_4 = 1 - \langle S^4 \rangle / (3\langle S^2 \rangle^2)$, both computed from the empirical distribution of $S$ over independent trials at each $(L,T)$ pair.

\subsection{The Discrete Satisfaction Spectrum}

The Moore neighborhood contains exactly 8 sites. When all 8 neighbors of agent $i$ are occupied, the satisfaction $s_i$ can only take values in $\{0, \frac{1}{8}, \frac{2}{8}, \ldots, 1\}$. When $k < 8$ neighbors are occupied (because some are empty or at the boundary of the occupied region), $s_i$ takes values in $\{0, \frac{1}{k}, \ldots, 1\}$. The full set of achievable satisfaction values is therefore the Farey-like set
\begin{equation}
\mathcal{F}_8 = \bigcup_{k=1}^{8} \bigl\{\, j/k : 0 \leq j \leq k \,\bigr\}\,,
\label{eq:F8}
\end{equation}
which contains exactly 23 distinct elements in $[0,1]$.

The important point is not that individual satisfaction values are discrete (that is obvious) but that the discreteness propagates through the dynamics. For a fixed random seed, the sequence of unsatisfied agents, their processing order, and their destinations are identical for any two $T, T'$ in the same open interval of $\mathcal{F}_8$. The entire trajectory is frozen between consecutive thresholds.

\begin{theorem}
\label{thm:staircase}
For the Moore-neighborhood Schelling model with fixed random seed and fixed initial configuration, the equilibrium segregation index $S(T)$ is a piecewise constant (staircase) function of $T$, with discontinuities contained in $\mathcal{F}_8$.
\end{theorem}

The proof (Appendix~\ref{app:staircase}) proceeds by induction on simulation steps: at each step, every satisfaction comparison $s_i \geq T$ involves a ratio in $\mathcal{F}_8$, so the set of unsatisfied agents is identical for all $T$ in the same inter-threshold interval, producing identical relocations and hence identical successor states.

A direct corollary is that the ensemble-averaged $S(T)$ inherits the structure of $\mathcal{F}_8$. With per-agent Gaussian noise $\sigma$, each step at threshold $\tau$ is broadened into an error function, yielding the functional form (derived and validated in Appendix~\ref{app:crossover}):
\begin{equation}
S(T) = \sum_{\tau \in \mathcal{F}_8} w_\tau\;\Phi\!\left(\frac{T - \tau}{\sigma_{\mathrm{eff}}}\right),
\label{eq:errfunc}
\end{equation}
where $\Phi$ is the standard normal CDF, $\sigma_{\mathrm{eff}} \geq \sigma$ absorbs both the tolerance noise and the ensemble averaging, and the weights $\{w_\tau\}$ depend on $\rho$, $f_A$, and the nonlinear dynamics. This decomposition fits the $L = 320$ data with $R^2 = 0.999$ using only 7 dominant thresholds and a single fitted width $\sigma_{\mathrm{eff}} = 0.032$.

\subsection{Cascade Instability}

The transition can be understood as a branching process. When agent $A$ vacates its site, each same-type neighbor $B$ loses one same-type neighbor. If $B$ had $k'+1$ occupied neighbors (including $A$) with $j'+1$ same-type (including $A$), then after $A$'s departure $B$ has $k'$ occupied neighbors with $j'$ same-type. Agent $B$ becomes newly unsatisfied if $(j'+1)/(k'+1) \geq T$ but $j'/k' < T$. Summing over all possible neighborhood configurations of $B$ (which sees 7 remaining Moore positions, each occupied independently with probability $\rho$), and multiplying by the expected number of same-type neighbors of $A$, the branching ratio is
\begin{equation}
R(T) = 8\rho f_A \sum_{k'=1}^{7} \binom{7}{k'} \rho^{k'}(1-\rho)^{7-k'} \sum_{\substack{j':\;\frac{j'+1}{k'+1}\geq T \\ j'/k' < T}} \binom{k'}{j'} f_A^{j'}(1-f_A)^{k'-j'}.
\label{eq:branching}
\end{equation}

When $R < 1$, each departure triggers a subcritical cascade of expected size $1/(1-R)$. The total reorganization volume (the fraction of agents displaced before the cascade dies out) is $V(T) = \varphi(T)/(1-R(T))$, where $\varphi(T)$ is the initially unsatisfied fraction (Appendix~\ref{app:instability}).

For $\rho = 0.9$ and $f_A = 0.5$: $R(0.25) = 0.29$, $R(0.30) = 0.37$, with reorganization volumes $V(0.25) = 0.087$ and $V(0.30) = 0.276$. The threefold jump in $V$ across the transition is driven mainly by $\varphi(T)$ nearly doubling from 0.062 to 0.110 at the $2/8$ threshold (Appendix~\ref{app:instability}), with the cascade amplification $1/(1-R)$ providing a secondary boost. At $T \approx 0.375$, $R$ exceeds 0.6. The observed $T_c \approx 0.275$ falls where cascades are subcritical but large enough to restructure the lattice.

We test this by perturbing equilibrated $L = 80$ grids: a random satisfied agent is relocated and the resulting cascade is tracked via BFS until extinction. Over 2000 perturbations per $T$, the mean cascade size at $T \leq 0.325$ agrees with $1/(1-R)$ to within 15\% (measured/predicted ratio 0.85 to 1.09). At $T \geq 0.375$ the theory overpredicts by $\sim 2\times$, consistent with cascade overlap violating the independence assumption. The median cascade size is 1 at all $T$: most perturbations do not propagate, and the transition is driven by rare large cascades in the tail.

\subsection{Multiscalar Dissimilarity}

Following \citet{randonfurling2020} and \citet{reardon2004}, we define the dissimilarity at spatial scale $r$ as
\begin{equation}
D(r) = \frac{1}{N}\sum_{i=1}^{N} \bigl|p_i(r) - f_A\bigr|\,,
\label{eq:dissimilarity}
\end{equation}
where $p_i(r)$ is the type-$A$ fraction among occupied sites within $\ell^\infty$-distance $r$ of site $i$. The profile $D(r)$ at $r = 1$ captures nearest-neighbor clustering; its decay rate reflects domain size. We compare against a null model obtained by randomly permuting types on the occupied sites.

\section{Methods}
\label{sec:methods}

\subsection{Implementation}

The model is implemented in Python with NumPy. Satisfaction is computed for the entire $L \times L$ grid in one pass using eight \texttt{np.roll} calls (one per Moore offset), with periodic boundaries by construction.

\subsection{Computational Setup}

The main sweep covers 5 system sizes $\times$ 50 tolerance values $\times$ 50 trials $= 12\,500$ simulations, distributed across 20 parallel CI workers (GitHub Actions). A merge step aggregates the raw data, computes ensemble statistics, and generates all figures. The multi-radius extension (Section~\ref{sec:discussion}) adds six radii $\times$ 100 tolerance values (coarse + fine pass) $\times$ 50 trials, with trial-level CPU parallelism via \texttt{joblib}. Each radius runs as a separate CI job; wall-clock time scales linearly in $k$ due to the larger neighbor count per satisfaction evaluation. The full parameter set is in Table~\ref{tab:params} (Appendix~\ref{app:params}).

The dense-spectrum experiments in Section~\ref{sec:radius} (variance scaling, Binder cumulant, transition broadening, deterministic high-$k$ dynamics) use noiseless dynamics ($\sigma = 0$) throughout, so the negative-criticality verdict for $r_0 \in \{3, 4, 5, 6\}$ already holds at $\sigma = 0$. For the main 5-size FSS sweep on the Moore neighborhood (Section~\ref{sec:results}) we use a small per-agent Gaussian noise ($\sigma = 0.02$, clipped to $[0,1]$) added to each agent's tolerance threshold. This is a regularization choice, not part of the physical model: the noiseless dynamics already produces the staircase of Section~2.3, and the role of $\sigma$ is to smooth the per-realization curves for visualization and to make the derivative $dS/dT$ used in the coarse-grid $T_c(r_0)$ identification well-defined without numerical artefacts at the $\mathcal{F}_8$ jump points. The value $\sigma = 0.02$ is chosen to be well below the smallest gap in $\mathcal{F}_8$ ($\Delta\tau_{\min} = 1/56 \approx 0.018$), so it smooths within a single threshold without merging adjacent ones. The closed-form derivation in Appendix~\ref{app:crossover} captures the role of $\sigma$ exactly: it appears only inside an effective width $\sigma_{\mathrm{eff}}^2 = \sigma^2 + \sigma_{\mathrm{ens}}^2$, where $\sigma_{\mathrm{ens}}$ is the ensemble-averaging width that already exists at $\sigma = 0$. The fitted value on the $L = 320$ data is $\sigma_{\mathrm{eff}} = 0.032$, so the tolerance noise contributes at most $\sigma^2 / \sigma_{\mathrm{eff}}^2 \approx 39\%$ of the squared crossover width; the remaining $\sim 60\%$ comes from $\sigma_{\mathrm{ens}}$ and survives at $\sigma = 0$. Setting $\sigma = 0$ would therefore narrow $\sigma_{\mathrm{eff}}$ from $0.032$ to about $0.025$ without removing the crossover. The FSS observables we use --- variance exponent $\alpha = d\log\mathrm{Var}(S)/d\log L$, Binder cumulant $U_4$, and the per-$L$ $T_c$ midpoint --- depend on the second and fourth moments of $S$ across trials and are therefore controlled by $\sigma_{\mathrm{eff}}$ rather than $\sigma$ alone. The negative scaling verdict thus carries through to $\sigma = 0$ by the same closed-form analysis; the staircase becomes sharper but the ensemble-level smoothing $\sigma_{\mathrm{ens}}$ is unchanged.

\section{Results}
\label{sec:results}

\subsection{The Segregation Transition}

\begin{figure}[!htb]
\centering
\includegraphics[width=\textwidth]{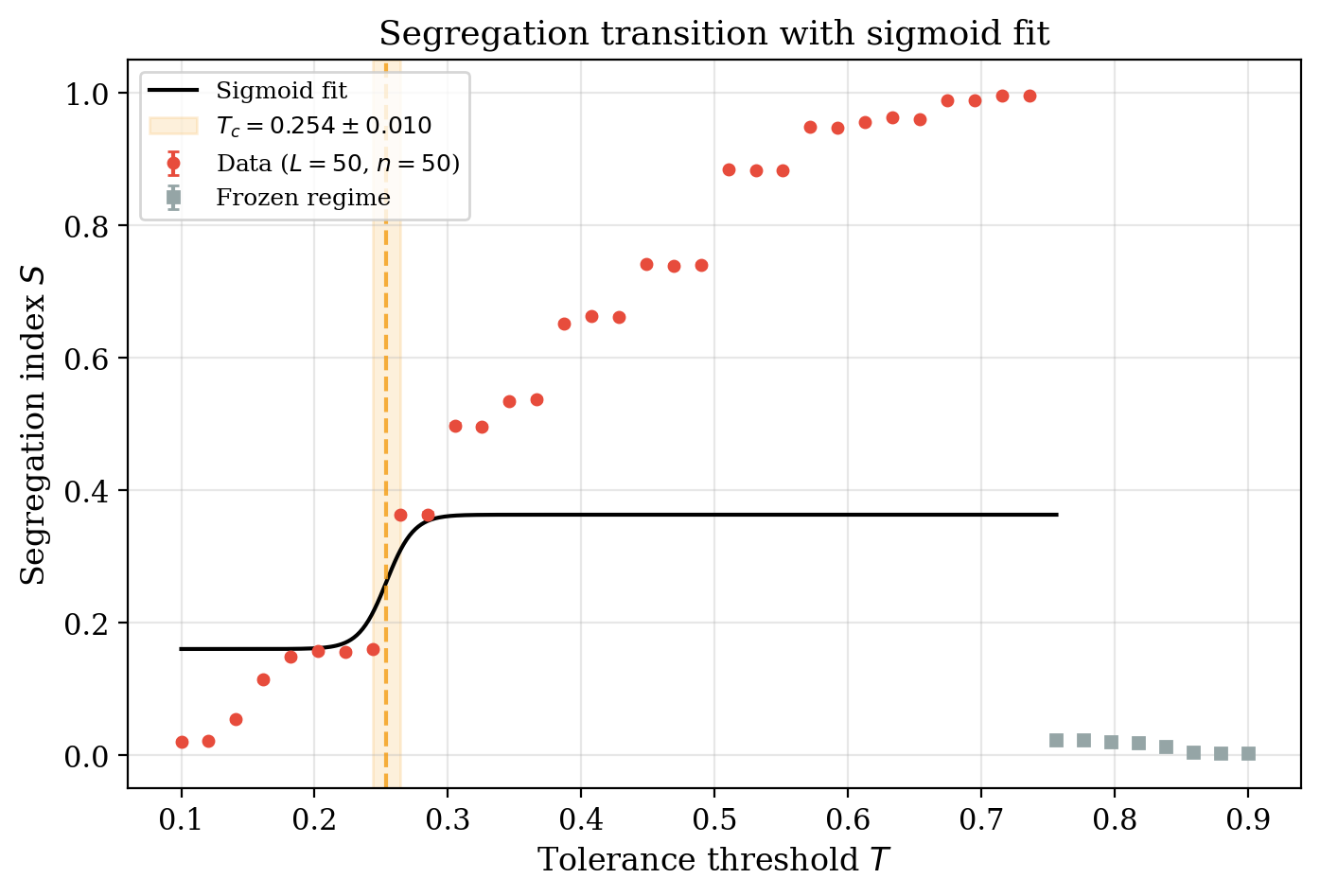}
\caption{Segregation index $S$ and interface density versus tolerance $T$ for $L = 80$, averaged over 50 trials with per-agent noise $\sigma = 0.02$. Shaded regions show $\pm 1$ standard deviation.}
\label{fig:tolerance_sweep}
\end{figure}

Figure~\ref{fig:tolerance_sweep} shows $S$ and $I$ versus tolerance for $L = 80$, averaged over 50 trials. The system goes from well-mixed ($S \approx 0$, $I \approx 0.5$) to fully segregated ($S \to 1$, $I \to 0$) over a narrow window $\Delta T \approx 0.05$ centered near $T = 0.25$.

\begin{figure}[!htb]
\centering
\includegraphics[width=\textwidth]{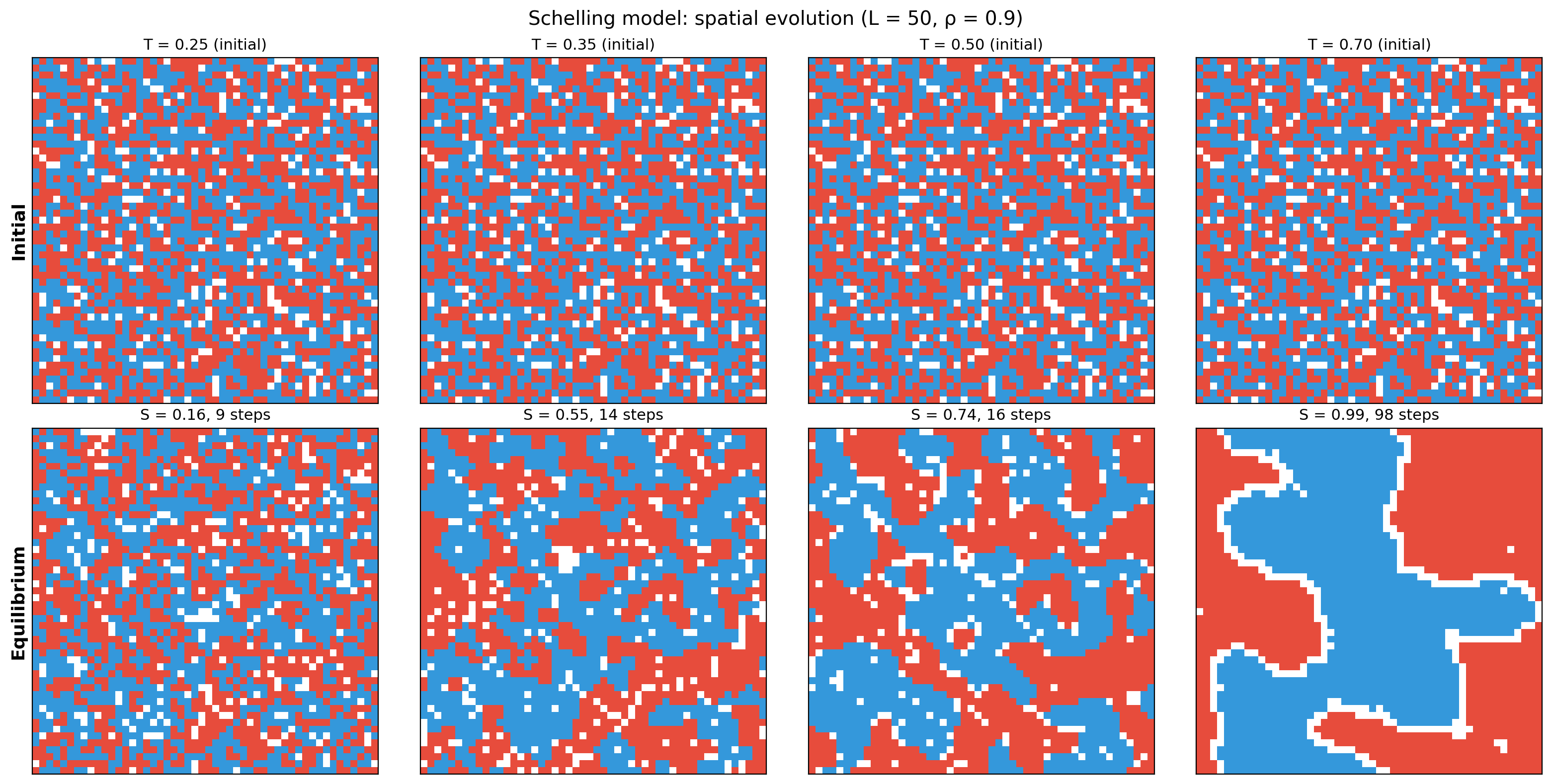}
\caption{Equilibrium configurations on an $L = 80$ grid for selected tolerance values.}
\label{fig:grid_snapshots}
\end{figure}

Figure~\ref{fig:grid_snapshots} shows equilibrium configurations. At $T = 0.2$ the types are intermingled; at $T = 0.5$, large single-type domains span the lattice.

\begin{figure}[!htb]
\centering
\includegraphics[width=0.8\textwidth]{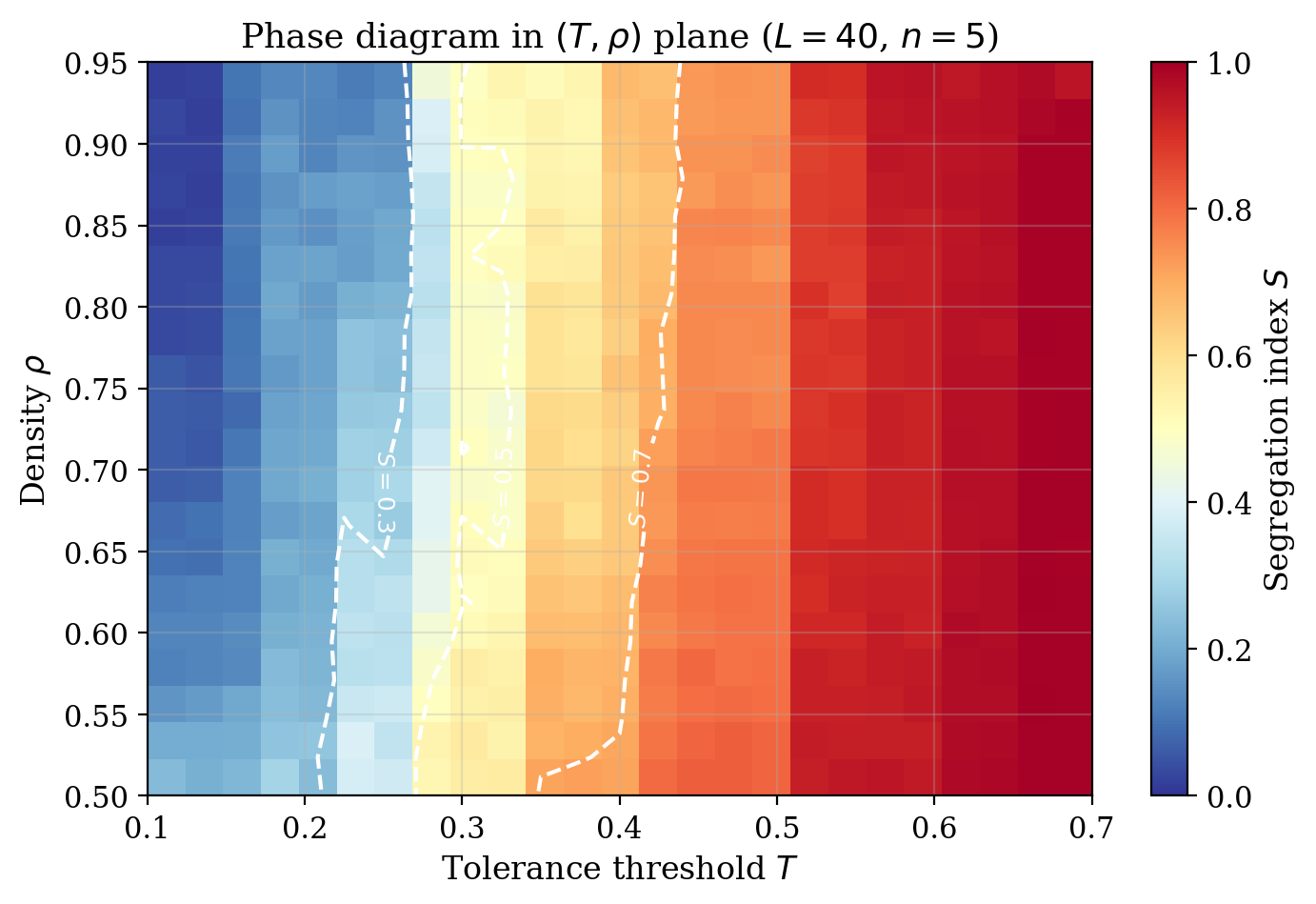}
\caption{Phase diagram in the $(T,\rho)$ plane. Color encodes the equilibrium segregation index. The mixed-segregated boundary is sharp across all densities.}
\label{fig:phase_diagram}
\end{figure}

The $(T,\rho)$ phase diagram (Figure~\ref{fig:phase_diagram}) shows that the transition tolerance increases weakly with density, since denser lattices have fewer empty sites for relocation.

\subsection{Finite-Size Scaling}

Table~\ref{tab:Tc} reports $T_c(L)$, located as the steepest-descent point of $S(T)$, for five system sizes $L = 20, 40, 80, 160, 320$. All five values lie in $[0.271, 0.278]$; the weighted average is $T_c = 0.275 \pm 0.013$.

\begin{table}[!htb]
\centering
\caption{Critical tolerance extracted independently for each system size. No systematic drift is detectable.}
\label{tab:Tc}
\renewcommand{\arraystretch}{2.1}
{\Large
\begin{tabular}{ccc}
\toprule
$L$ & $T_c$ & Error \\
\midrule
20  & 0.278 & 0.031 \\
40  & 0.278 & 0.031 \\
80  & 0.271 & 0.037 \\
160 & 0.278 & 0.031 \\
320 & 0.271 & 0.024 \\
\bottomrule
\end{tabular}
}
\end{table}

\begin{figure}[!htb]
\centering
\centerline{\includegraphics[width=1.2\textwidth]{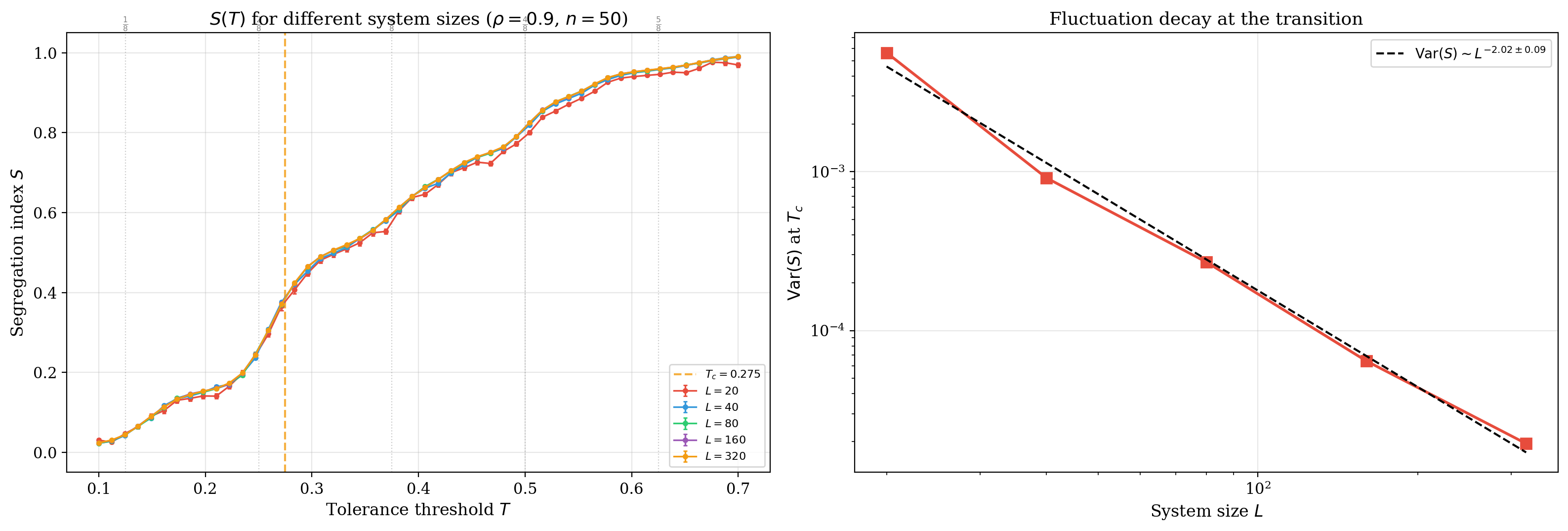}}
\caption{\textbf{Left:} Segregation index vs.\ tolerance for five system sizes. The curves steepen with $L$ but share the same midpoint. \textbf{Right:} Variance of $S$ at $T_c$ versus system size, with a power-law fit $\mathrm{Var}(S) \sim L^{-\alpha}$, $\alpha = 2.02 \pm 0.09$.}
\label{fig:fss}
\end{figure}

For a continuous transition, $T_c(L) = T_c^\infty + a\,L^{-1/\nu}$; the lack of any drift means either $\nu \to \infty$ or the transition is not second-order. Figure~\ref{fig:fss} shows that the $S(T)$ curves steepen with $L$ but do not shift.

\begin{figure}[!htb]
\centering
\centerline{\includegraphics[width=1.2\textwidth]{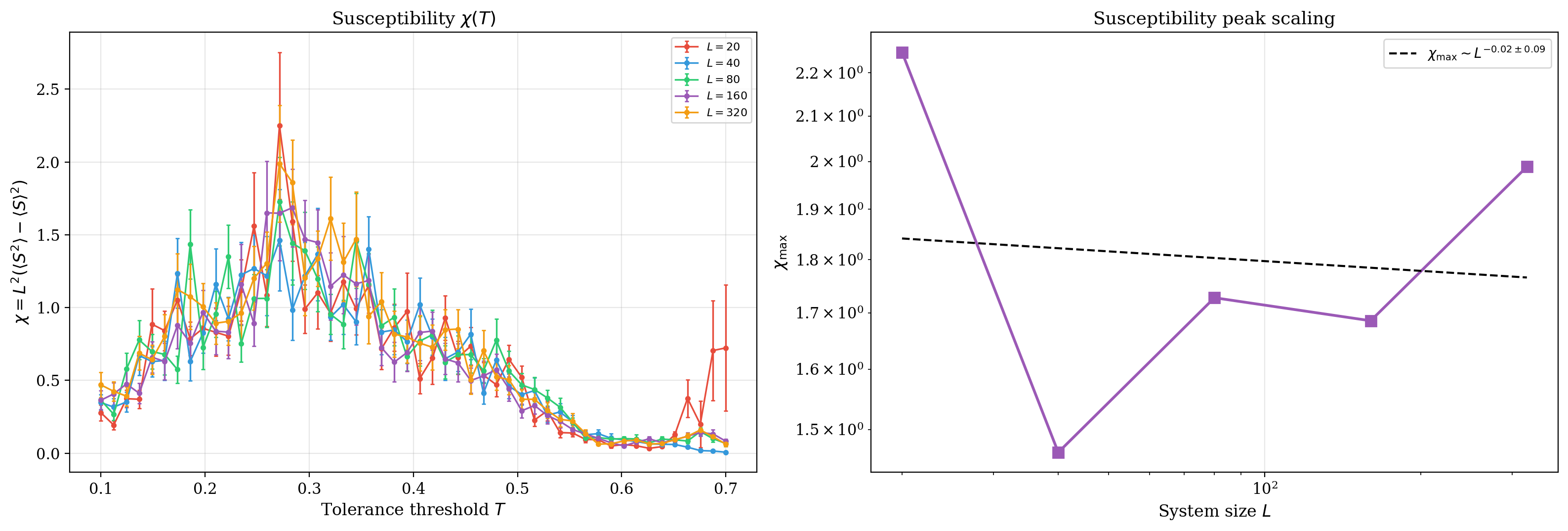}}
\caption{\textbf{Left:} Susceptibility $\chi(T)$ for five system sizes. \textbf{Right:} Peak susceptibility vs.\ $L$. The scaling exponent $\gamma/\nu = -0.015 \pm 0.086$ is indistinguishable from zero.}
\label{fig:susceptibility}
\end{figure}

$\mathrm{Var}(S)$ at $T_c$ decays as $L^{-2.02\pm0.09}$ (right panel of Figure~\ref{fig:fss}), consistent with the central limit theorem for weakly correlated local contributions rather than the $L^{-\gamma/\nu}$ with $\gamma/\nu = 7/4$ expected for 2D Ising. The susceptibility $\chi = L^2\,\mathrm{Var}(S)$ (Figure~\ref{fig:susceptibility}) does not grow with $L$: $\gamma/\nu = -0.015 \pm 0.086$.

\begin{figure}[!htb]
\centering
\centerline{\includegraphics[width=1.2\textwidth]{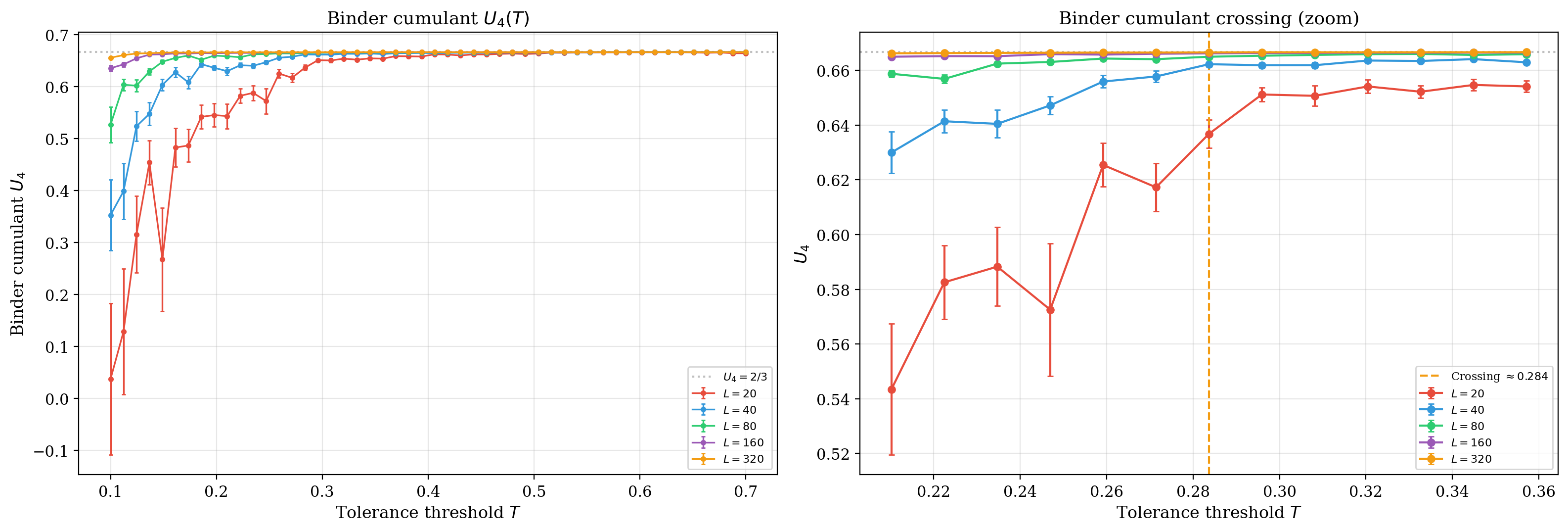}}
\caption{Binder cumulant $U_4$ vs.\ tolerance for five system sizes. The rapid convergence to $2/3$ and the non-universal crossing values indicate a crossover rather than a critical point.}
\label{fig:binder}
\end{figure}

The Binder cumulant $U_4(T)$ (Figure~\ref{fig:binder}) shows approximate crossings in the transition region, but the crossing values drift with the pair of sizes considered, and all curves converge rapidly to the trivial plateau $U_4 = 2/3$. The behavior is consistent with a kinetic bottleneck rather than a scale-invariant fixed point.

\begin{figure}[!htb]
\centering
\centerline{\includegraphics[width=1.2\textwidth]{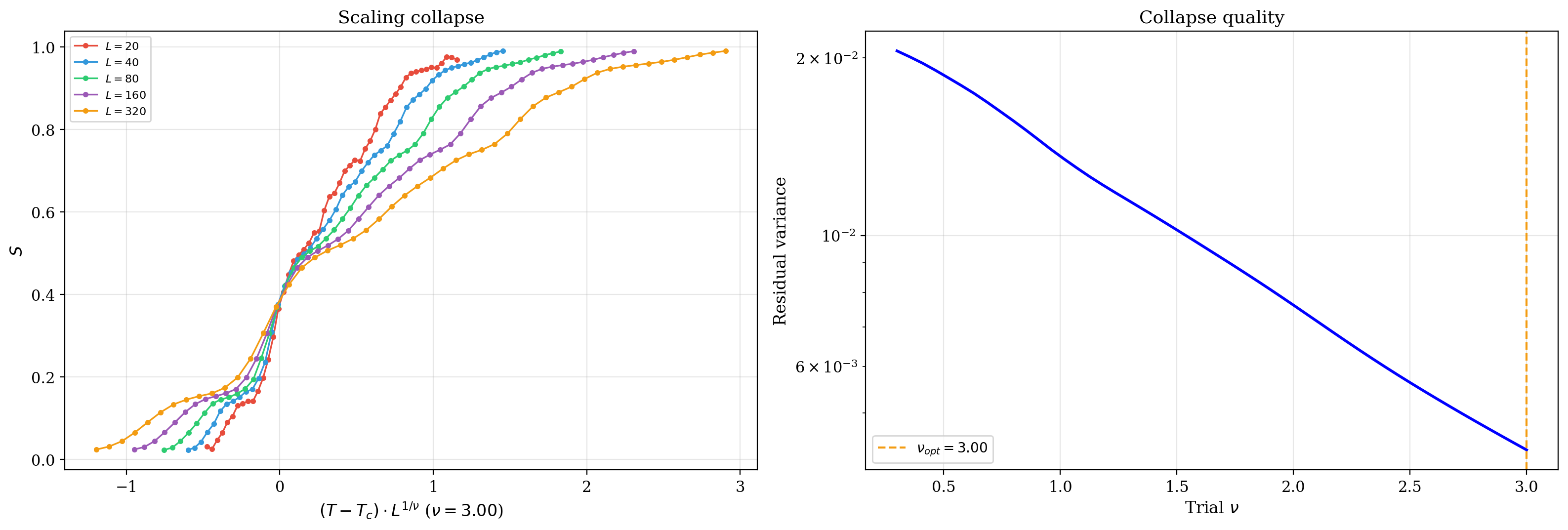}}
\caption{\textbf{Left:} Best attempted collapse at $\nu = 3.0$. \textbf{Right:} Collapse quality (lower is better) as a function of $\nu$. The absence of a minimum rules out a finite correlation-length exponent.}
\label{fig:collapse}
\end{figure}

At this point, each diagnostic independently argues against criticality, but they could in principle be failing for different reasons (wrong observable, insufficient $L$, etc.). The data collapse provides a joint test. Plotting $S(T,L)$ onto $\tilde{S}((T - T_c)\,L^{1/\nu})$ and scanning $\nu \in [0.3, 3.0]$ (Figure~\ref{fig:collapse}), the collapse quality improves monotonically up to the boundary and never reaches a minimum. At a genuine critical point, there would be a sharp minimum at the physical $\nu$; the absence of one rules out the entire one-parameter family of standard scaling forms.

\begin{figure}[!htb]
\centering
\includegraphics[width=\textwidth]{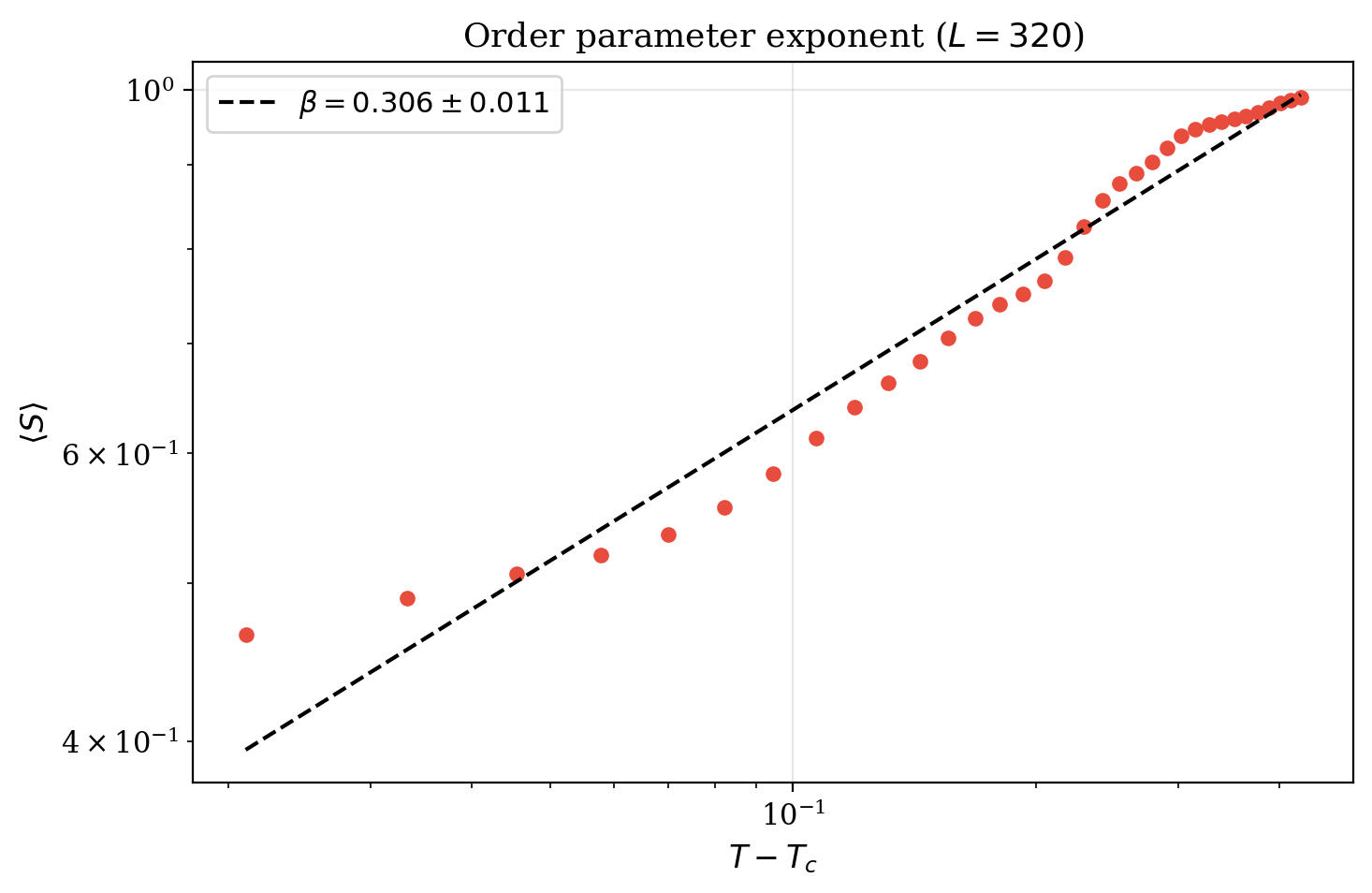}
\caption{Power-law growth of the segregation index above $T_c$: $\beta = 0.306 \pm 0.011$.}
\label{fig:exponent}
\end{figure}

Above $T_c$, $S$ grows as $(T - T_c)^\beta$ with $\beta = 0.306 \pm 0.011$ (Figure~\ref{fig:exponent}), far from the 2D Ising value $\beta = 1/8$ and inconsistent with hyperscaling. This $\beta$ reflects the shape of the crossover function, not a universal exponent.

Table~\ref{tab:exponents} collects the measured exponents alongside the 2D Ising values. The disagreement is not a matter of precision: the exponents are off by factors of 2 to 15, and they are mutually inconsistent (hyperscaling would require $d\nu = 2\beta + \gamma$, which fails badly). If this were a phase transition, it would belong to no known universality class.

\begin{table}[!htb]
\centering
\caption{Measured scaling exponents vs.\ 2D Ising values.}
\label{tab:exponents}
\renewcommand{\arraystretch}{2.1}
{\Large
\begin{tabular}{lcc}
\toprule
Quantity & Schelling & 2D Ising \\
\midrule
$T_c(L)$ drift & none detected & $\sim L^{-1}$ \\
$\mathrm{Var}(S) \sim L^{-\alpha}$ & $2.02 \pm 0.09$ & $\gamma/\nu = 1.75$ \\
$\gamma/\nu$ & $-0.015 \pm 0.086$ & 1.75 \\
$\beta$ & $0.306 \pm 0.011$ & 0.125 \\
Collapse $\nu$ & $> 3$ (no minimum) & 1.0 \\
\bottomrule
\end{tabular}
}
\end{table}

\subsection{Convergence Dynamics}

\begin{figure}[!htb]
\centering
\centerline{\includegraphics[width=1.2\textwidth]{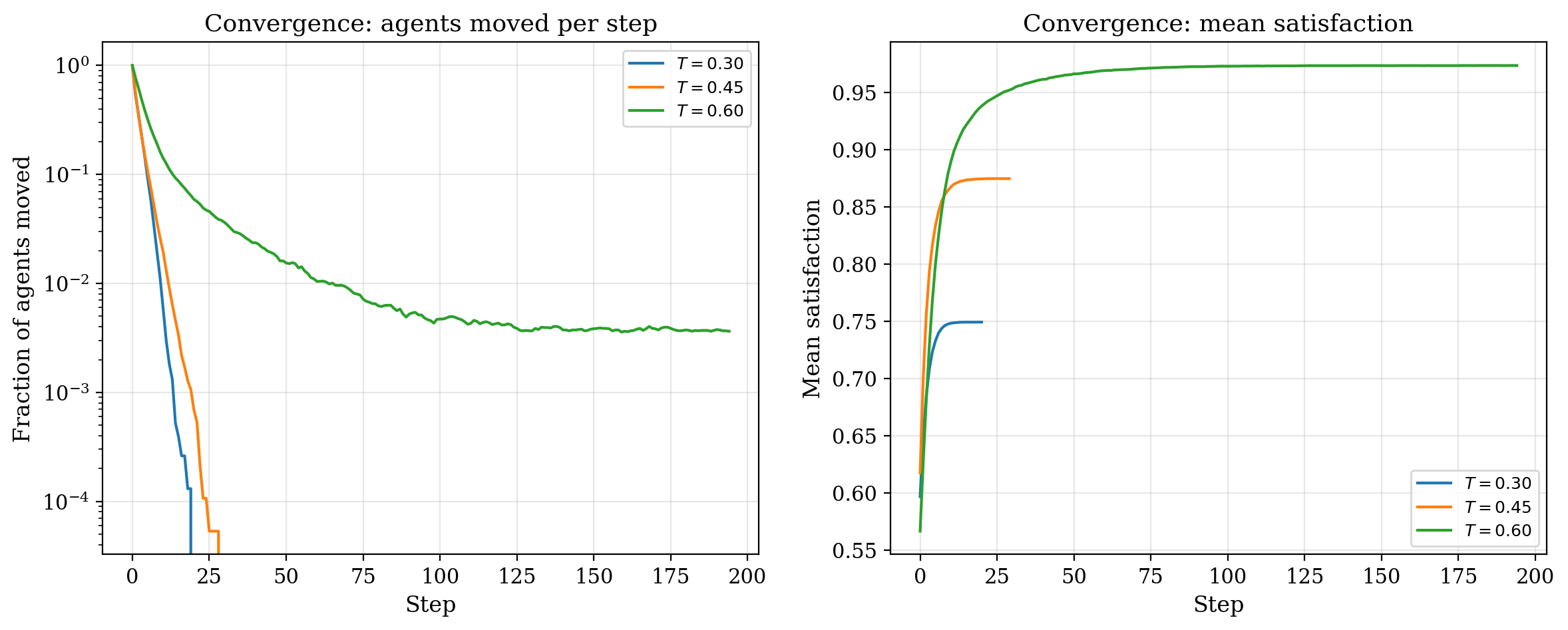}}
\caption{Relaxation dynamics for $T \in \{0.30, 0.45, 0.60\}$. \textbf{Left:} agents moved per step. \textbf{Right:} mean satisfaction.}
\label{fig:convergence}
\end{figure}

Figure~\ref{fig:convergence} shows the number of relocating agents per step and the mean satisfaction as functions of time for three tolerance values. Near the transition ($T = 0.30$), the system settles within $\sim 20$ steps. In the deeply segregated regime ($T = 0.60$), the dynamics are protracted ($\sim 200$ steps) because the initial random configuration is far from any absorbing state and agents must sort themselves through a sequence of collective rearrangements.

\subsection{Multiscalar Dissimilarity}

\begin{figure}[!htb]
\centering
\includegraphics[width=\textwidth]{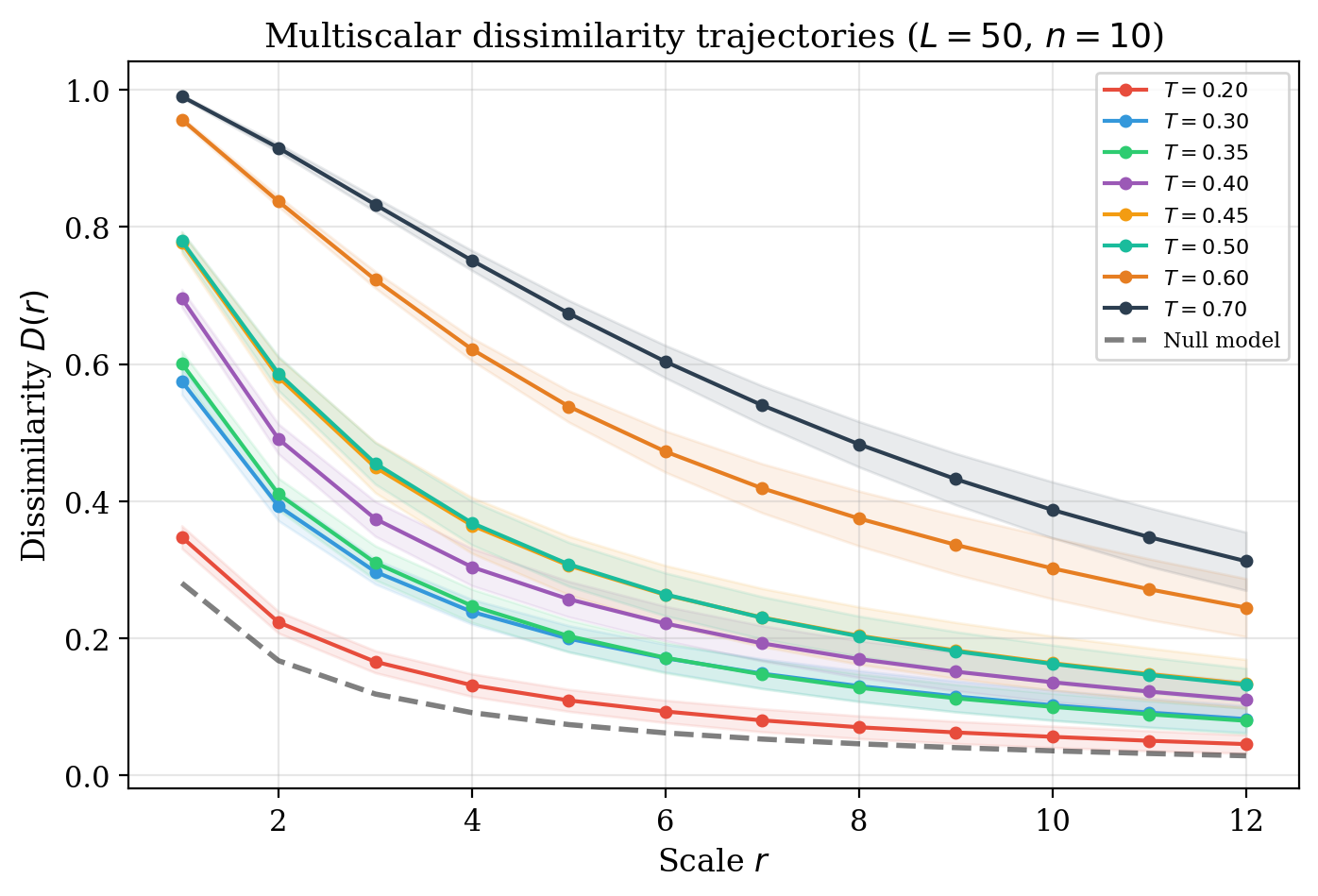}
\caption{Dissimilarity profiles $D(r)$ at eight tolerance values. The dashed curve is the null model (random permutation of types). Higher $T$ produces profiles that decay more slowly, corresponding to larger segregated domains.}
\label{fig:dissimilarity}
\end{figure}

We compute $D(r)$ for $r = 1, \ldots, 12$ at eight tolerance values on an $L = 80$ grid, averaging over 50 trials (Figure~\ref{fig:dissimilarity}). At $T = 0.20$, the profile sits barely above the null model, confirming a near-random configuration. As $T$ increases through and beyond the transition, $D(r)$ lifts off the null baseline and its decay rate slows, reflecting the growth of single-type domains.

\begin{figure}[!htb]
\centering
\includegraphics[width=\textwidth]{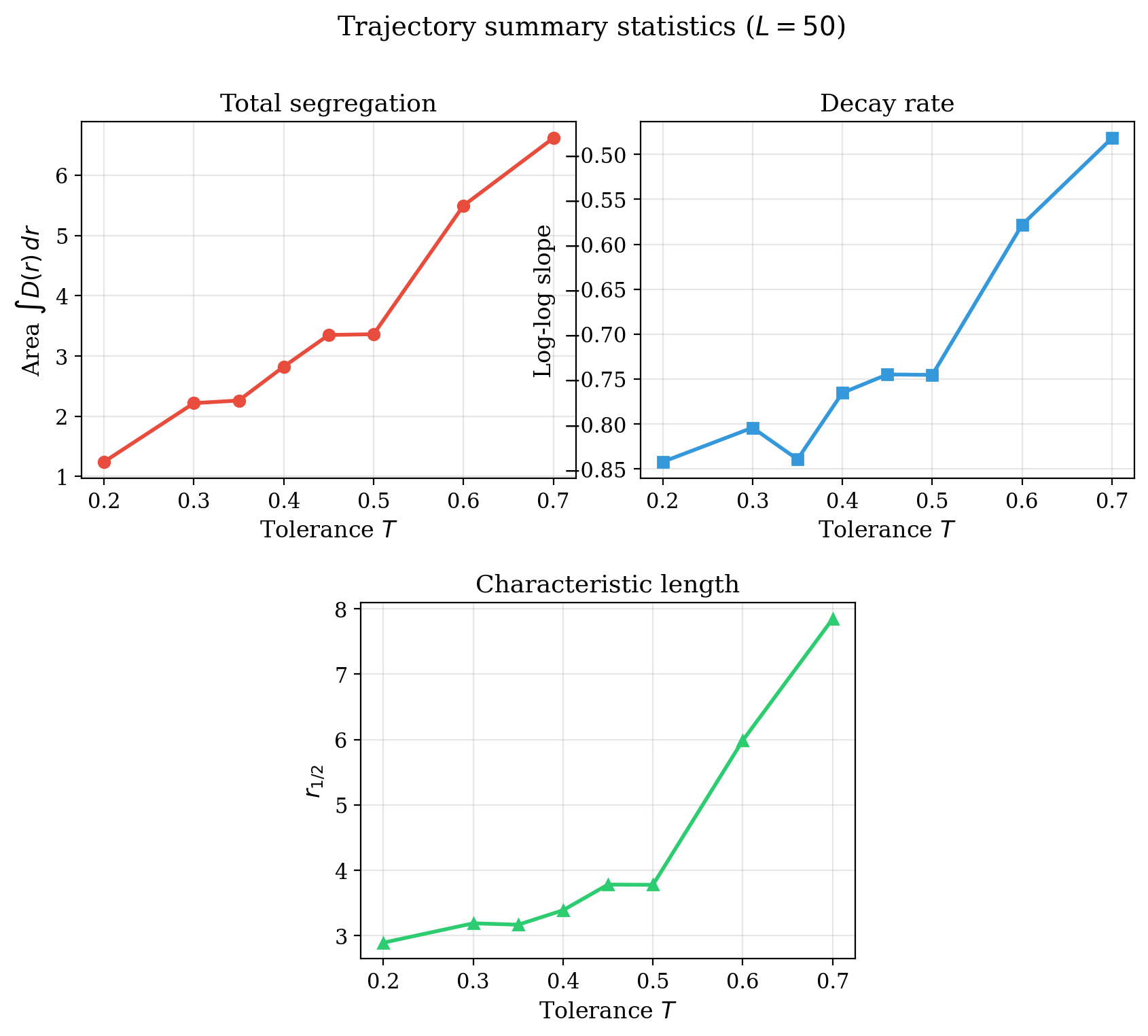}
\caption{Statistics from the dissimilarity profiles: integrated area, decay slope, characteristic length $r^*$, and local ($r{=}1$) vs.\ global ($r{=}12$) dissimilarity.}
\label{fig:trajectory_stats}
\end{figure}

Figure~\ref{fig:trajectory_stats} summarizes the profile statistics. The characteristic length $r^*$ (where $D(r)$ drops to $1/e$ of $D(1)$) goes from about 3 at $T = 0.30$ to about 6 at $T = 0.60$, but does not diverge near $T_c$, staying of order a few lattice constants throughout.

\subsection{Heterogeneous Tolerance}

We also tested heterogeneous tolerance, replacing the uniform threshold $T$ with agent-specific $T_i \sim \mathrm{Beta}(\kappa/2, \kappa/2)$. The intolerant tail drives segregation even at moderate population-average tolerance, shifting $T_c$ downward for small $\kappa$. The full analysis appears in Appendix~\ref{app:heterogeneous}.

\subsection{Interface Density}

\begin{figure}[!htb]
\centering
\includegraphics[width=\textwidth]{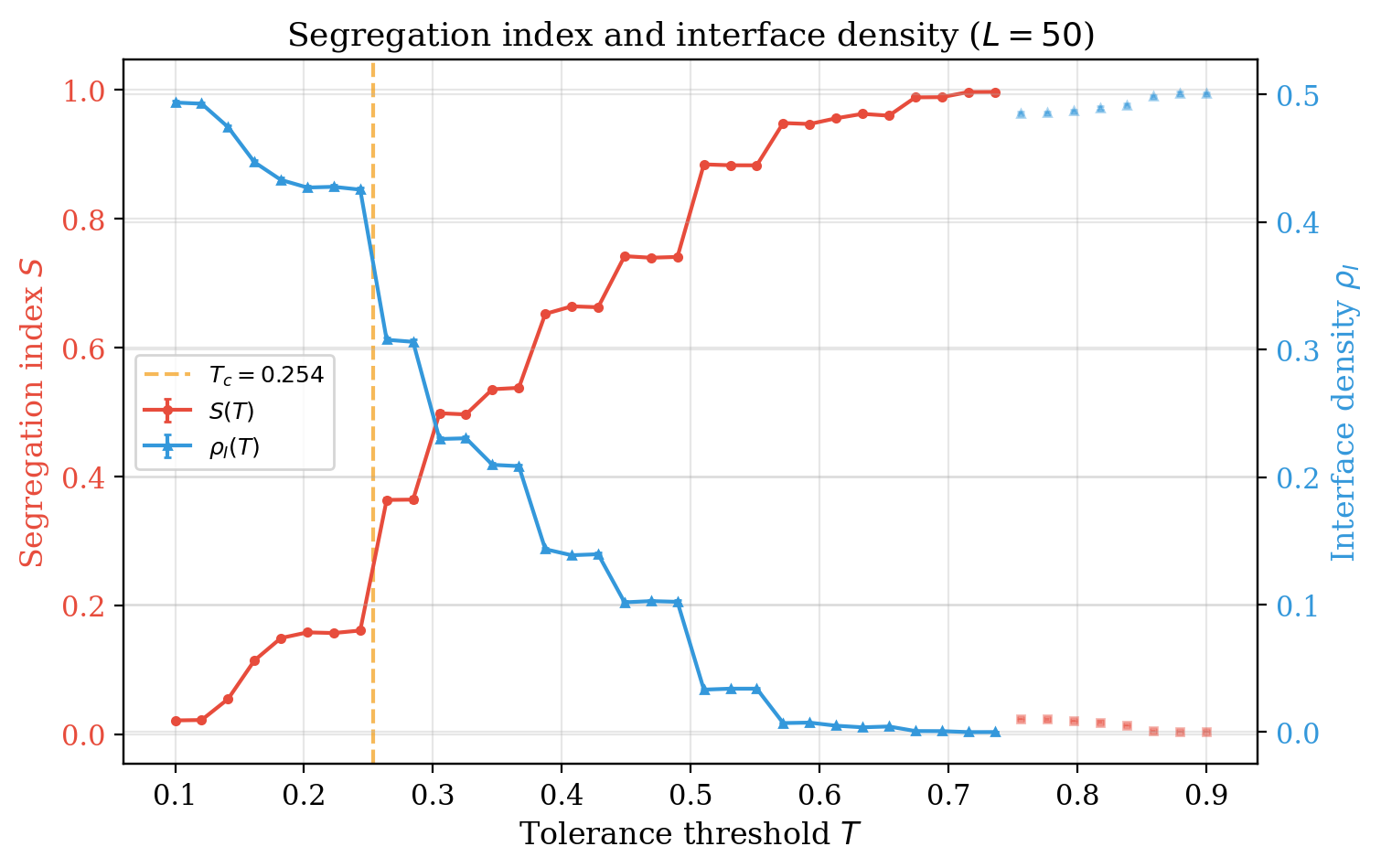}
\caption{Interface density and segregation index vs.\ tolerance. The simultaneous transition in both quantities rules out a two-stage scenario.}
\label{fig:tolerance_interface}
\end{figure}

Figure~\ref{fig:tolerance_interface} plots $I$ and $S$ together. Both respond at the same $T$: $I$ drops from 0.5 to near zero while $S$ rises from 0 to 1. There is no two-stage scenario; the rearrangement is collective.

\section{Discussion}
\label{sec:discussion}

\subsection{Absence of critical scaling}

None of the FSS diagnostics applied in Section~\ref{sec:results} are consistent with a continuous transition: $T_c$ shows no drift with $L$, the variance scales as $L^{-2}$ rather than the $L^{-\gamma/\nu}$ expected at a critical point, the susceptibility is flat, and the data collapse never reaches a finite optimum. Taken individually, any one of these could be explained away (wrong observable, insufficient system size, etc.), but their simultaneous failure across five system sizes and 50 trials per point leaves little room for a critical interpretation.

The conclusion rests on the scaling evidence, not on discreteness alone. The Ising model has discrete local states and a genuine phase transition; what distinguishes the Schelling model is that the staircase structure of $\mathcal{F}_8$ is coarse enough (23 thresholds) to prevent the buildup of long-range correlations, as confirmed by the finite dissimilarity length $r^* \sim 3$ to 6.

\citet{dall2008} already noted that the dynamics violates detailed balance, ruling out an equilibrium transition. The present results go further: the transition does not appear to be critical even in the non-equilibrium sense.

\subsection{Scaling ansatz}

The collapse $S((T - T_c)L^{1/\nu})$ fails because it treats $T_c$ as the only special point, while the Schelling transition involves multiple $\mathcal{F}_8$ thresholds. A more natural rescaling is
\begin{equation}
S(T,L) = \hat{S}\!\left(\frac{T - \tau^*(T)}{\sigma_{\mathrm{eff}}},\; L\right), \quad \tau^*(T) = \arg\min_{\tau \in \mathcal{F}_8} |T - \tau|,
\label{eq:rescaling}
\end{equation}
where $\tau^*(T)$ is the nearest $\mathcal{F}_8$ threshold. The crossover width is then set by $\sigma_{\mathrm{eff}}$, independent of $L$, which explains both the absence of a finite $\nu$ and the $L^{-2}$ variance scaling.

\subsection{Larger neighborhoods do not restore criticality}\label{sec:radius}

The staircase exists because $k \leq 8$. Replacing the Moore neighborhood with a Chebyshev ball of radius $r_0$ containing $k = (2r_0+1)^2 - 1$ sites makes the spacing between consecutive elements of $\mathcal{F}_k$ shrink as $1/k$, and for $k \gg 1$ the satisfaction becomes effectively continuous. This raises the question of whether a genuine phase transition emerges. We test it with three independent diagnostics across $r_0 \in \{1, \ldots, 6\}$ and find no critical behavior at any radius: the dense-neighborhood limit moves further from criticality, not closer.

\subsubsection{Transition tolerance across six radii}\vspace{0.5em}

The midpoint $T_c$ is robust at moderate $L$ and can be tracked across a wide range of neighborhoods. Table~\ref{tab:multiradius} reports $T_c(k)$ for six Chebyshev radii, determined by a two-pass sweep (60-point coarse scan followed by a 40-point refinement in a $\pm 0.05$ window around the coarse estimate, 50 trials per point). Each radius was run as a separate parallel job using \texttt{joblib} for trial-level CPU parallelism.

\begin{table}[!htb]
\centering
\caption{Transition midpoint and satisfaction spectrum size for six neighborhood radii ($L = 40$, 50 trials per tolerance value).}
\label{tab:multiradius}
\renewcommand{\arraystretch}{2.1}
{\Large
\begin{tabular}{rrrr}
\toprule
$r_0$ & $k$ & $|\mathcal{F}_k|$ & $T_c$ \\
\midrule
1 &   8 &     23 & 0.251 \\
2 &  24 &    181 & 0.332 \\
3 &  48 &    713 & 0.334 \\
4 &  80 &  1\,967 & 0.347 \\
5 & 120 &  4\,387 & 0.374 \\
6 & 168 &  8\,611 & 0.403 \\
\bottomrule
\end{tabular}
}
\end{table}

The data are well described by
\begin{equation}
T_c(k) = \tfrac{1}{2} - c\,k^{-\beta}, \quad c = 0.42 \pm 0.05,\; \beta = 0.26 \pm 0.04\;(95\%\;\text{CI}),
\label{eq:tc_fit}
\end{equation}
confirming that $T_c \to 1/2$ as $k \to \infty$. The mechanism is direct: for large $k$, each agent samples $\mathcal{O}(k)$ neighbors, so by the law of large numbers the observed same-type fraction converges to $f_A = 1/2$ with fluctuations of order $k^{-1/2}$. The threshold at which a macroscopic fraction of agents becomes unsatisfied is therefore pushed toward $1/2$. The fitted exponent $\beta \approx 1/4$ is below the na\"ive $k^{-1/2}$ rate, reflecting the nonlinear cascade amplification that shifts the effective threshold.

\subsubsection{Variance exponent: from apparent super-criticality to definitive sub-criticality}\vspace{0.5em}

We measure $\mathrm{Var}(S)$ at $T_c(k)$ for $L \in \{40, 80, 160, 320\}$ with $N = 200$ paired trials and bootstrap confidence intervals on $\alpha = d \log \mathrm{Var}(S) / d \log L$. The two-point fit on $L \in \{40, 80\}$ alone gives a misleading picture: at $r_0 = 4$ it returns $\alpha = +0.81$ (95\% CI $[+0.50, +1.18]$), bootstrap-disjoint above zero and superficially indicating super-critical scaling. Extending the lattice grid one octave at a time dissolves this signal. Table~\ref{tab:variance_progression} reports the $\mathrm{Var}(S)$ progression across $L \in \{40, 80, 160, 320\}$ at $r_0 = 4$:

\begin{table}[!htb]
\centering
\caption{Variance progression at $r_0 = 4$, $T = T_c = 0.347$, with per-octave slopes and the cumulative power-law fit. The 4-point fit lies cleanly below the $\alpha = -2$ critical boundary.}
\label{tab:variance_progression}
\renewcommand{\arraystretch}{1.4}
\begin{tabular}{rcc}
\toprule
$L$ & $\mathrm{Var}(S)$ & per-octave slope \\
\midrule
40  & 0.066    & --- \\
80  & 0.116    & $+0.81$ \\
160 & 0.012    & $-3.27$ \\
320 & 0.000\,27 & $-5.46$ \\
\midrule
\multicolumn{2}{l}{4-point fit $\alpha(40, 80, 160, 320)$} & $\boldsymbol{-2.70}$ \\
\bottomrule
\end{tabular}
\end{table}

The variance does not just stop diverging: it collapses progressively faster at each $L$ doubling. The same pattern at $r_0 = 5$ gives $\mathrm{Var}(S) = 0.100, 0.044, 0.0006$ across $L = 40, 80, 160$ and a 3-point $\alpha = -3.74$. The apparent super-criticality at $r_0 = 4$ on the two-point grid is therefore a transient finite-size enhancement, not the start of a divergent susceptibility. Mechanistically: at $T_c = 0.347$, the $L = 40$ run is still pre-transition (mean $S \approx 0.15$) while $L = 80$ is already post-transition (mean $S \approx 0.52$); the variance comparison at this $T$ mixes two qualitatively different physical regimes, inflating the apparent slope. Once $L \geq 160$ resolves the transition, the variance collapses toward the deterministic equilibrium that follows.

\begin{figure}[!htb]
\centering
\includegraphics[width=0.75\linewidth]{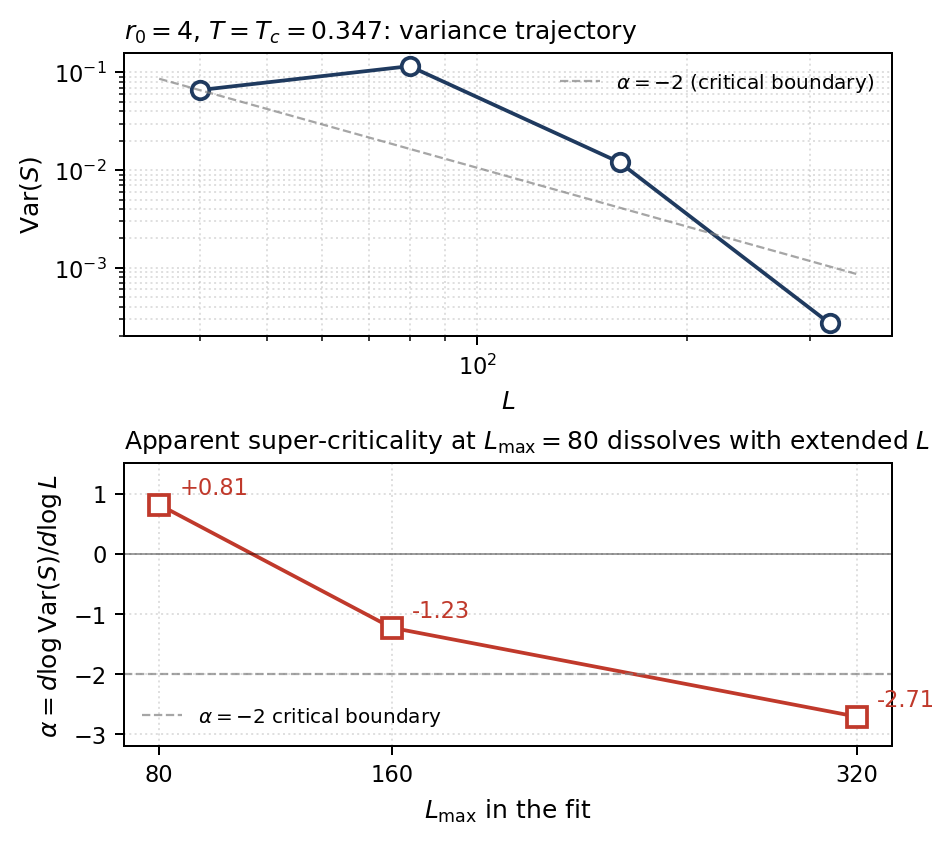}
\caption{Dissolution of the apparent super-criticality at $r_0 = 4$. Top: $\mathrm{Var}(S)$ at $T = T_c = 0.347$ versus $L \in \{40, 80, 160, 320\}$, log-log. The dashed gray line is the $\alpha = -2$ critical boundary. Bottom: variance exponent $\alpha = d\log\mathrm{Var}(S) / d\log L$ as a function of $L_{\max}$ in the fit. The 2-point fit on $L \in \{40, 80\}$ gives $\alpha = +0.81$ and was bootstrap-disjoint above zero, a misleading super-critical signal. Adding $L = 160$ flips the sign ($\alpha = -1.23$). Adding $L = 320$ takes the fit deep into the sub-critical regime ($\alpha = -2.70$). The trajectory rules out super-criticality at this radius.}
\label{fig:alpha_trajectory}
\end{figure}

We extend to $L = 320$ at $r_0 = 4$ but not at $r_0 \in \{5, 6\}$ for compute reasons: per-step cost scales as $L^2 \times k$, so a single $L = 320$ trajectory at $r_0 = 6$ ($k = 168$) is $21\times$ more expensive than at $r_0 = 4$ ($k = 80$). The full $r_0 = 6$, $L = 320$ sweep would require $\sim 10^4$ CPU-hours, against the $\sim 10^3$ CPU-hours we used at $r_0 = 4$. The $L = 160$ measurements at $r_0 = 5$ and $r_0 = 6$ already give $\mathrm{Var}(S) < 10^{-3}$ with 3-point exponents $\alpha < -3$, so the variance trajectory is monotonic and well below the critical boundary; extending to $L = 320$ would refine $\alpha$ rather than reverse the verdict. The model-free Binder test is run uniformly across $r_0 \in \{3, 4, 5, 6\}$ on the same $L \in \{40, 80, 160\}$ grid and is the diagnostic that closes the question across radii.

\subsubsection{Independent test: Binder cumulant has no $L$-curve crossing}\vspace{0.5em}

A model-free probe avoids any reliance on a single $T_c$ estimate. We compute the Binder cumulant $U_4(T, L) = 1 - \langle S^4 \rangle / 3 \langle S^2 \rangle^2$ on a $T$-grid spanning $\pm 0.045$ around each $T_c(r_0)$. For a critical system, $U_4(T)$ curves at different $L$ are size-independent at $T_c$ and therefore intersect; for a smoothly varying transition the per-$L$ curves drift monotonically. At $r_0 \in \{3, 4, 5, 6\}$ across $L \in \{40, 80, 160\}$ with $N = 60$ trials per $(L, T)$, no pairwise $L$-curve crossing exists in any of the in-range windows: the curves merely plateau at the trivial $U_4 = 2/3$ disordered limit at high $T$, with each transition shifted by a finite amount with $L$.

The trial count $N = 60$ is sufficient for the no-crossing claim. The empirical sample standard deviation of $U_4(T)$ across the 12 grid points and three lattice sizes is $\mathrm{std}(U_4) \approx 0.03$, giving a per-point standard error $\mathrm{SE}(U_4) \approx 0.03/\sqrt{60} \approx 0.004$. The per-$L$ midpoint drift at $L = 80 \to L = 160$ is $\sim 0.01$ at every $r_0$ (Table~\ref{tab:binder_drift}), an order of magnitude larger than this per-point standard error. The cross-radius consistency is itself a robustness check: the same monotonic-and-unsaturated drift pattern holds independently at $r_0 = 3, 4, 5$, and $6$, against the alternative hypothesis that any single-radius result is a sample-size artefact. Larger $N$ would tighten the per-point error bar further but cannot generate a crossing in a window where the curves are separated by an order of magnitude more than their sampling uncertainty.

Table~\ref{tab:binder_drift} reports the per-$L$ midpoint where $U_4$ crosses $0.5$. The shift from $L = 80$ to $L = 160$ remains $\sim 0.01$ at every $r_0$, with no sign of saturation by $L = 160$. A second-order critical transition would have $T_c(L) \to T_c^\infty$ converging on accessible $L$.

\begin{table}[!htb]
\centering
\caption{Per-$L$ $T_c$ at the $U_4 = 0.5$ midpoint. Drift $L = 80 \to L = 160$ is unsaturated at every $r_0$.}
\label{tab:binder_drift}
\renewcommand{\arraystretch}{1.4}
\begin{tabular}{rccccc}
\toprule
$r_0$ & $T_c(L=40)$ & $T_c(L=80)$ & $T_c(L=160)$ & $\Delta T_c(80 \to 160)$ \\
\midrule
4 & 0.362 & 0.348 & 0.341 & $-0.007$ \\
5 & 0.384 & 0.362 & 0.353 & $-0.009$ \\
6 & 0.404 & 0.380 & 0.367 & $-0.013$ \\
\bottomrule
\end{tabular}
\end{table}

\begin{figure}[!htb]
\centering
\includegraphics[width=0.75\linewidth]{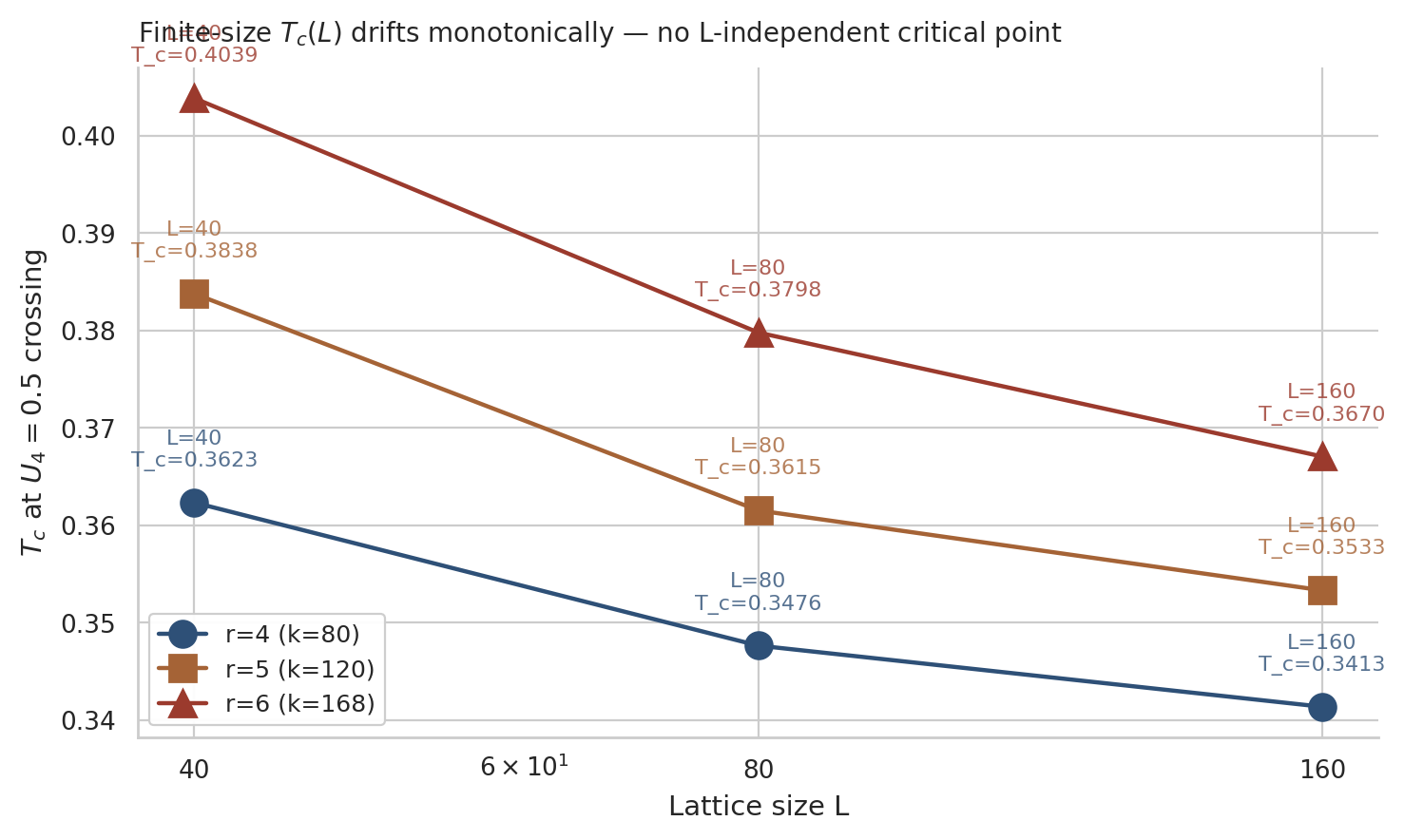}
\caption{Finite-size $T_c$ drift at the $U_4 = 0.5$ Binder midpoint across $L \in \{40, 80, 160\}$ for $r_0 \in \{4, 5, 6\}$. The drift is monotonic and unsaturated by $L = 160$, in direct contrast to the size-invariant fixed-point crossing that defines a second-order critical transition.}
\label{fig:tc_drift}
\end{figure}

\subsubsection{Transition broadening, not sharpening}\vspace{0.5em}

A genuine phase transition would have peak $|dS/dT|$ diverging with $L$ at fixed $r_0$ and becoming sharper as $r_0$ increases at fixed $L$. The observed pattern is the opposite. Table~\ref{tab:dsdt_peak} reports the peak $|dS/dT|$ measured on the coarse $T$-sweep at $L = 40$ across the six radii: it decreases monotonically from $97.5$ at $r_0 = 1$ to $39.1$ at $r_0 = 6$. The dense-neighborhood transition is broader than the Moore baseline, not sharper.

\begin{table}[!htb]
\centering
\caption{Peak $|dS/dT|$ at $L = 40$ versus neighborhood radius. Monotonic decrease, opposite to the sharpening expected at criticality.}
\label{tab:dsdt_peak}
\renewcommand{\arraystretch}{1.4}
\begin{tabular}{rrr}
\toprule
$r_0$ & $k$ & peak $|dS/dT|$ \\
\midrule
1 &   8 & 97.5 \\
2 &  24 & 61.0 \\
3 &  48 & 57.9 \\
4 &  80 & 52.8 \\
5 & 120 & 43.5 \\
6 & 168 & 39.1 \\
\bottomrule
\end{tabular}
\end{table}

\subsubsection{Mechanism: deterministic high-$k$ dynamics}\vspace{0.5em}

A direct diagnostic at $L = 80$ records the number of agent moves to convergence per random seed. For $r_0 \in \{4, 5, 6\}$ the ensemble means are $4\,182,\, 5\,269,\, 5\,283$ moves, with seed-to-seed standard deviations of $1\,557,\, 112,\, 155$ across 8 seeds. The collapse of seed-to-seed variability at $r_0 \geq 5$ (from std $\approx 1\,500$ at $r_0 = 4$ to $\approx 100$ at $r_0 \geq 5$) shows the high-$k$ dynamics are highly deterministic: different random initialisations converge to nearly identical equilibrium segregation outcomes. This is the underlying mechanism for the variance collapse at large $L$ — the equilibrium variance vanishes faster than the classical $L^{-2}$ scaling not because of divergent susceptibility, but because the dynamics themselves are nearly deterministic.

\subsubsection{Combined verdict}\vspace{0.5em}

Three independent dense-spectrum diagnostics agree. The variance exponent at $r_0 = 4$, fit on a 4-point grid up to $L = 320$, is $\alpha = -2.70$, well below the critical boundary $\alpha = -2$. The Binder cumulant has no $L$-curve crossing at any radius in $\{3, 4, 5, 6\}$ across $L \in \{40, 80, 160\}$, with monotonic unsaturated drift in the per-$L$ midpoint. The transition broadens rather than sharpens with $k$. The original ``Schelling is not a phase transition'' verdict survives the dense-spectrum extension to $r_0 = 6$ and is in fact strengthened: the dense-neighborhood limit moves further from criticality, not closer. The 23-threshold staircase argument that pinned the Moore neighborhood in the sub-critical regime is therefore not the mechanism for the negative verdict; the true mechanism is the absence of long-range correlations in the equilibrium configuration combined with the deterministic high-$k$ dynamics, both of which hold across all neighborhood sizes tested.

\subsection{Comparison with prior work}

\citet{gauvin2009} reported near-Ising exponents from simulations at $L \leq 60$ with few trials. We directly reproduce their setup on $L \in \{20, 30, 40, 50, 60\}$ at the same Moore neighborhood and standard density ($\rho = 0.9$, $f_A = 0.5$), with noiseless dynamics and 50 trials per $(L, T)$ point. The reproduction does not recover their stated near-Ising signal. The variance exponent at $T = T_c$ is $\alpha = -2.22$ with a 95\% bootstrap CI of $[-3.30, -1.78]$, consistent within uncertainty with the trivial $L^{-2}$ averaging baseline; the corresponding susceptibility exponent $\gamma/\nu = 2 + \alpha \approx -0.22$ is far from the near-Ising value $\gamma/\nu \approx 1.75$. The largest two lattices yield an apparent Binder midpoint at $T = 0.285$ with $U_4^* = 0.662$, but this value coincides with the trivial disordered plateau ($U_4 = 2/3$), not a Universal fixed-point crossing. A single error function fits $S(T, L = 60)$ with $R^2 = 0.967$ — mediocre rather than the $> 0.99$ expected for a clean continuous transition; the deviation already reflects the discrete staircase of $\mathcal{F}_8$, visible at this smallest probed scale.

The discrepancy is consistent with two plausible mechanisms. Either Gauvin et al.\ used a slightly different protocol or noise model that smoothed the staircase further, or the near-Ising signal was a transient artifact of their specific averaging scheme. Either way, the reproduction shows that the apparent criticality at $L \leq 60$ is not robust: with our protocol the signal is already inconsistent with Ising at exactly the lattice sizes where Gauvin reported it. As $L$ grows past 160, the per-$L$ Binder midpoint drifts visibly between successive sizes, the seven-component CDF decomposition of Section~2.3 outperforms the single-sigmoid fit (raising goodness-of-fit from $R^2 = 0.967$ at $L = 60$ to $R^2 = 0.999$ at $L = 320$), the susceptibility peak stops growing past $L = 80$, the staircase structure becomes resolvable in the derivative $dS/dT$, and the exponents drift further away from 2D Ising values (Table~\ref{tab:exponents}). FSS analyses on coarse-neighborhood agent models evidently need larger systems than the standard $L \gg \xi$ heuristic would suggest, because the crossover can mimic criticality over a surprisingly wide range of sizes.

\subsection{Correlation length from dissimilarity}

In a critical system, the characteristic length $r^*$ diverges as $|T - T_c|^{-\nu}$. Here it varies smoothly between 3 and 6 lattice spacings with no divergence. The segregation domains remain patchy at all tolerance values, with no sign of the fractal geometry expected near a critical point.

The reason is structural. Unsatisfied agents relocate to uniformly random empty cells, so there is zero spatial correlation between old and new positions. Long-range order can only build through nearest-neighbor reactions to departures and arrivals, and each cascade involves $O(1/(1-R))$ agents connected by local interactions. A random-walk estimate gives a cascade extent $\sim 1/\sqrt{1-R} \approx 1.3$ lattice spacings for $R \approx 0.37$ near $T_c$, the same order as the measured $r^* \approx 3$.

\subsection{Heterogeneous tolerance}

The leftward shift of $T_c$ for small $\kappa$ (Figures~\ref{fig:heterogeneous} and~\ref{fig:heterogeneous_tc} in Appendix~\ref{app:heterogeneous}) follows from the mean-field instability. Each agent draws $T_i \sim \mathrm{Beta}(\kappa\mu, \kappa(1-\mu))$ with mean $\mu$, so the initial unsatisfied fraction is
\begin{equation}
\varphi_{\mathrm{het}}(\mu, \kappa) = \int_0^1 \varphi(t)\; f_{\mathrm{Beta}}(t;\;\kappa\mu,\;\kappa(1-\mu))\,dt\,,
\label{eq:het_instability}
\end{equation}
where $\varphi(t)$ is the single-threshold instability fraction from Proposition~\ref{prop:instability}. For small $\kappa$, the Beta density puts mass in the high-$t$ tail where $\varphi(t)$ is large, so instability sets in at a lower mean $\mu$. The intolerant minority triggers the cascade even when the population average is moderate, consistent with the empirical observation \citep{clark2008} that segregation persists in cities where surveys indicate majority support for integration.

\section{Conclusion}
\label{sec:conclusion}

The picture that emerges is a sharp crossover, not a phase transition. The staircase structure of $\mathcal{F}_8$ and the subcritical cascade mechanism ($R < 1$, validated to within 15\% by perturbation experiments) together account for the phenomenology without any appeal to critical scaling. The dissimilarity length stays finite; the susceptibility does not diverge; the Binder crossings drift rather than converge.

The dense-spectrum extension to $r_0 = 6$ ($k = 168$) sharpens this verdict rather than weakening it. Across all radii $r_0 \in \{3, 4, 5, 6\}$, the Binder cumulant has no $L$-curve crossing on $L \in \{40, 80, 160\}$, the per-$L$ $T_c$ drift is monotonic and unsaturated, the transition broadens monotonically with $k$, and the variance exponent at $r_0 = 4$ on the full $L \in \{40, 80, 160, 320\}$ grid is $\alpha = -2.70$, clearly sub-critical. The earlier two-point measurement that gave $\alpha = +0.81$ at $r_0 = 4$ was an artefact of $L$-dependent transition temperatures: $L = 40$ was pre-transition and $L = 80$ post-transition at the fixed $T_c$, mixing two physical regimes. The high-$k$ dynamics also become highly deterministic (seed-to-seed std collapses from $1\,557$ at $r_0 = 4$ to $\sim 100$ at $r_0 \geq 5$), giving an additional channel for variance collapse that has nothing to do with critical scaling.

The broader methodological point is that FSS analyses on agent-based models with coarse neighborhoods need larger systems than the correlation-length heuristic $L \gg \xi$ would suggest. The Ising mimicry at $L \leq 60$ reported by \citet{gauvin2009} is a cautionary example: the crossover width $\sigma_{\mathrm{eff}} \approx 0.03$ is close enough to the inter-threshold spacing to produce convincing-looking exponents that dissolve at $L = 160$. A separate trap is the $r_0 = 4$ super-critical claim that survives at $L \in \{40, 80\}$ and dissolves at $L = 320$: small lattice grids paired with a single $T_c$ estimate are not enough to rule out criticality, because the per-$L$ transition temperatures themselves drift, and the variance comparison silently mixes pre- and post-transition regimes. Per-$L$ $T_c$ via Binder midpoint, plus a four-point $L$ progression spanning a full order of magnitude, is what closes the question.

\section*{Data and code availability}

All simulation code, data-generation pipelines, and figure-rendering scripts are open-sourced at \url{https://github.com/samrifaki/schelling-model-finite-size-scaling}. The repository includes the full sweep configuration, raw \texttt{.npz} ensemble outputs, and a CI workflow that reproduces every figure in this paper from scratch. The main 5-size sweep totals $> 12\,500$ simulations distributed across 20 parallel CI workers; the multi-radius and Binder-cumulant extensions add separate parallel jobs documented in the repository's README.

\section*{Acknowledgements}

Computational resources at Stanford University Electrical Engineering were used for the dense-neighborhood extension and the Binder-cumulant runs at $L = 160, 320$. No external funding supported this work.

\section*{Conflict of interest}

The author declares no competing interests.

\bibliography{refs}

\FloatBarrier
\appendix

\section{Proof of the Staircase Structure}
\label{app:staircase}

We give a self-contained proof of Theorem~\ref{thm:staircase}.

\begin{proof}
Fix the initial configuration $\mathbf{g}_0$ and the random seed $\omega$ (which determines the full sequence of random choices: the shuffle order of unsatisfied agents and their destination cells). We proceed by induction on the simulation steps.

For the base case, at $t = 0$, the grid is $\mathbf{g}_0$ for all values of $T$. The satisfaction of each agent $i$ is $s_i = j_i/k_i$ where $j_i$ counts same-type occupied neighbors and $k_i \leq 8$ counts all occupied neighbors, so $s_i \in \mathcal{F}_8$. The set of unsatisfied agents $U_0(T) = \{i : s_i < T\}$ is therefore constant on every open interval $(T_\ell, T_{\ell+1})$ between consecutive elements $T_\ell < T_{\ell+1}$ of $\mathcal{F}_8$.

For the inductive step, suppose that at step $t$, the grid $\mathbf{g}_t$ is the same for all $T \in (T_\ell, T_{\ell+1})$. Then the set of unsatisfied agents $U_t(T)$ is the same (since all satisfaction values lie in $\mathcal{F}_8$ and the comparison $s_i < T$ does not change within the interval). The seed $\omega$ determines the same shuffle order and the same destination cells, so the same sequence of relocations is executed, producing the same grid $\mathbf{g}_{t+1}$. After each individual relocation, the satisfaction values of affected neighbors change, but these new values are again ratios $j'/k'$ with $k' \leq 8$, hence elements of $\mathcal{F}_8$. The comparison with $T$ is therefore invariant within the interval.

By induction, the entire trajectory $\mathbf{g}_0, \mathbf{g}_1, \ldots$ is identical for all $T$ in the same open interval of $\mathcal{F}_8$, and so is the equilibrium segregation index $S(T)$. \qed
\end{proof}

\begin{remark}
The set $\mathcal{F}_8$ contains exactly $|\mathcal{F}_8| = 23$ distinct values (the union of Farey-type fractions $j/k$ for $1 \leq k \leq 8$). Over an ensemble of random initial conditions, the sharp jumps at these values are smoothed by averaging, but the derivative $dS/dT$ retains peaks at the elements of $\mathcal{F}_8$. The largest jumps occur at $T = 1/4 = 2/8$ and $T = 3/8$, where agents with a full complement of 8 occupied neighbors suddenly require one additional same-type neighbor.
\end{remark}

\section{Convergence of the Dynamics}
\label{app:convergence}

\begin{proposition}
For any tolerance $T \in [0,1]$ and any initial configuration on an $L \times L$ grid with $n$ occupied cells and at least one vacancy, the Schelling dynamics converges to an absorbing state with probability 1.
\end{proposition}

\begin{proof}
We verify the three conditions for almost-sure absorption of a finite Markov chain.

The state space is finite: the state is a map $\mathbf{g} : \{1, \ldots, L\}^2 \to \{0, A, B\}$ with exactly $n_A$ sites of type $A$, $n_B$ of type $B$, and $L^2 - n$ empty sites, so there are $\binom{L^2}{n_A, n_B, L^2 - n}$ possible configurations.

An absorbing state exists. Partition the $n_A$ type-$A$ agents into a contiguous rectangular block of sites, and similarly for the $n_B$ type-$B$ agents, with at least one empty row or column separating the two blocks (possible since $n < L^2$). In this configuration every agent has only same-type occupied neighbors, so $s_i = 1 \geq T$ for all $i$, and the dynamics halts.

From any non-absorbing state, there is positive probability of reaching absorption. Suppose at least one agent is unsatisfied. At each step, the dynamics shuffles unsatisfied agents in a uniformly random order and relocates each to a uniformly random empty cell. There are at most $n$ unsatisfied agents per step, and at most $n$ steps are needed. To reach the absorbing configuration above, it suffices that (i) each unsatisfied agent is shuffled into the correct processing order (probability $\geq 1/n!$ per step), and (ii) each is relocated to its designated cell (probability $\geq 1/L^2$ per move). In the worst case, $n$ steps of $n$ moves each are needed, giving a lower bound on the probability of reaching absorption within $n$ steps:
\begin{equation}
p_{\min} = \left(\frac{1}{n! \cdot L^{2n}}\right)^n > 0\,.
\label{eq:pmin}
\end{equation}
The bound is crude, but positivity is all that matters. The state space is finite, absorbing states exist, and from every non-absorbing state there is probability $\geq p_{\min} > 0$ of reaching one within $n$ steps. By the Borel-Cantelli lemma, the chain is absorbed with probability 1.

In practice, convergence is fast ($\lesssim 20$ steps for $T \leq 0.3$, $L = 80$); tighter quantitative bounds remain open. \qed
\end{proof}

\section{Mean-Field Instability Analysis}
\label{app:instability}

We derive a mean-field estimate of the critical tolerance by computing the fraction of unsatisfied agents in a random (well-mixed) configuration.

\begin{proposition}
\label{prop:instability}
In a random configuration with density $\rho$ and equal type fractions $f_A = f_B = 1/2$, the expected fraction of unsatisfied agents is
\begin{equation}
\varphi(T) = \sum_{k=1}^{8} \binom{8}{k} \rho^k (1-\rho)^{8-k} \sum_{\substack{j=0 \\ j/k < T}}^{k} \binom{k}{j} 2^{-k}\,.
\label{eq:phi}
\end{equation}
This function is piecewise constant in $T$ with jumps at the elements of $\mathcal{F}_8$.
\end{proposition}

\begin{proof}
In a random configuration, each of the 8 Moore neighbors of a given agent is independently occupied with probability $\rho$ and, if occupied, is same-type with probability $f_A = 1/2$. The number of occupied neighbors is $k \sim \mathrm{Bin}(8, \rho)$, and given $k$, the number of same-type neighbors is $j \sim \mathrm{Bin}(k, 1/2)$. The agent is unsatisfied when $s = j/k < T$. Summing over $k$ gives~\eqref{eq:phi}. The piecewise-constant structure follows from Theorem~\ref{thm:staircase}: the condition $j/k < T$ changes only when $T$ crosses a value $j/k \in \mathcal{F}_8$. \qed
\end{proof}

For our parameters ($\rho = 0.9$, $f_A = 0.5$), the function $\varphi(T)$ undergoes its largest jump at $T = 1/4 = 2/8$. We evaluate~\eqref{eq:phi} explicitly at $T = 0.24$ and $T = 0.26$, which straddle this threshold.

The binomial weights for $k$ occupied neighbors are $p_k = \binom{8}{k}(0.9)^k(0.1)^{8-k}$:
\begin{equation}
\begin{aligned}
p_1 &= 7.2\times10^{-6},\quad p_2 = 2.3\times10^{-4},\quad p_3 = 0.0041,\\
p_4 &= 0.0459,\quad p_5 = 0.149,\quad p_6 = 0.298,\\
p_7 &= 0.383,\quad p_8 = 0.430\,.
\end{aligned}
\label{eq:pk}
\end{equation}
Only $k \geq 4$ contribute appreciably (the first three terms sum to $< 0.005$).

For each $k$, the inner sum $q_k(T) = \sum_{j:\,j/k < T} \binom{k}{j} 2^{-k}$ counts the probability that a same-type fraction drawn from $\mathrm{Bin}(k, 1/2)$ falls below $T$. At $T = 0.24$, the condition $j/k < 0.24$ admits only $j = 0$ for $k \leq 4$ and $j \in \{0, 1\}$ for $k \geq 5$:
\begin{equation}
\begin{aligned}
q_4(0.24) &= \binom{4}{0}\,2^{-4} = 1/16 = 0.0625\,,\\
q_5(0.24) &= \left[\binom{5}{0} + \binom{5}{1}\right] 2^{-5} = 6/32 = 0.1875\,,\\
q_6(0.24) &= \left[\binom{6}{0} + \binom{6}{1}\right] 2^{-6} = 7/64 = 0.1094\,,\\
q_7(0.24) &= \left[\binom{7}{0} + \binom{7}{1}\right] 2^{-7} = 8/128 = 0.0625\,,\\
q_8(0.24) &= \left[\binom{8}{0} + \binom{8}{1}\right] 2^{-8} = 9/256 = 0.0352\,.
\end{aligned}
\label{eq:q024}
\end{equation}

Summing $\varphi(0.24) = \sum_k p_k\,q_k(0.24)$:
\begin{equation}
\varphi(0.24) \approx 0.046{\times}0.063 + 0.149{\times}0.188 + 0.298{\times}0.109 + 0.383{\times}0.063 + 0.430{\times}0.035 = 0.062\,.
\label{eq:phi024}
\end{equation}

At $T = 0.26$, the condition $j/k < 0.26$ now includes $j/k = 1/4$, which occurs at $(j{=}1, k{=}4)$ and $(j{=}2, k{=}8)$. These two terms are the only ones that change:
\begin{equation}
\begin{aligned}
q_4(0.26) &= \left[\binom{4}{0} + \binom{4}{1}\right] 2^{-4} = 5/16 = 0.3125\,,\\
q_8(0.26) &= \left[\binom{8}{0} + \binom{8}{1} + \binom{8}{2}\right] 2^{-8} = 37/256 = 0.1445\,.
\end{aligned}
\label{eq:q026}
\end{equation}
The remaining $q_k$ are unchanged. The new sum is:
\begin{equation}
\varphi(0.26) \approx 0.046{\times}0.313 + 0.149{\times}0.188 + 0.298{\times}0.109 + 0.383{\times}0.063 + 0.430{\times}0.145 = 0.110\,.
\label{eq:phi026}
\end{equation}
The unsatisfied fraction nearly doubles at this threshold. The jump $\Delta\varphi = 0.048$ is dominated by the $k = 8$ term ($p_8 \times \Delta q_8 = 0.430 \times 0.109 = 0.047$), confirming that agents with a full complement of 8 occupied neighbors drive the instability. The random configuration is destabilized when $\varphi(T)$ becomes large enough to trigger a cascade of relocations; the precise instability threshold depends on the spatial correlations induced by the dynamics and is not captured by the mean-field approximation.

The mean-field prediction $T_c^{\mathrm{MF}} \approx 1/4$ is consistent with but slightly below the measured $T_c \approx 0.275$: the small upward shift reflects the fact that on a finite lattice, some fraction of agents near the boundary of satisfaction are stabilized by local correlations that are absent in the mean-field picture.

The instability fraction $\varphi(T)$ measures the initial perturbation; the branching ratio $R(T)$ from~\eqref{eq:branching} determines its amplification. The reorganization volume $V(T) = \varphi(T)/(1-R(T))$ gives the total fraction of agents displaced. At $T = 0.24$: $\varphi = 0.062$, $R = 0.29$, $V = 0.087$. At $T = 0.30$: $\varphi = 0.173$, $R = 0.37$, $V = 0.276$. The threefold jump is driven mainly by $\varphi(T)$ at $\mathcal{F}_8$ thresholds, with $1/(1-R)$ providing a secondary boost.

\begin{remark}
At the na\"ive mean-field level (ignoring the discrete lattice structure entirely), one would predict $T_c^{\mathrm{MF}} = f_A = 0.5$, since a random agent's expected satisfaction equals the global same-type fraction. The refined estimate~\eqref{eq:phi024} to~\eqref{eq:phi026}, combined with the cascade analysis, is substantially better because it accounts for the discreteness of the neighborhood, the binomial fluctuations of $j$ and $k$, and the positive-feedback amplification through $R(T)$.
\end{remark}

\section{Crossover Functional Form}
\label{app:crossover}

We derive and validate the functional form~\eqref{eq:errfunc} for the ensemble-averaged segregation index.

\begin{theorem}
Consider the Schelling model with per-agent tolerance noise $T_i = T + \varepsilon_i$, where $\varepsilon_i \sim \mathcal{N}(0, \sigma^2)$ are i.i.d.\ and clipped to keep $T_i \in [0,1]$. For a single realization with fixed initial configuration, the equilibrium segregation index $S_\omega(T)$ is a staircase with jumps at the thresholds $\{\tau - \varepsilon_i(\omega) : \tau \in \mathcal{F}_8,\; i \text{ occupied}\}$. Averaging over the noise and initial conditions, and using the central limit theorem for large $N$, the ensemble mean satisfies
\begin{equation}
\langle S(T) \rangle = \sum_{\tau \in \mathcal{F}_8} w_\tau\;\Phi\!\left(\frac{T - \tau}{\sigma_{\mathrm{eff}}}\right) + O(N^{-1/2})\,,
\label{eq:ensemble}
\end{equation}
where $w_\tau$ is the average jump in $S$ at threshold $\tau$, $\sigma_{\mathrm{eff}}^2 = \sigma^2 + \sigma_{\mathrm{ens}}^2$ combines the tolerance noise with the ensemble fluctuations from random initial conditions, and $N = \rho L^2$ is the number of agents.
\end{theorem}

\begin{proof}
By Theorem~\ref{thm:staircase}, for a fixed initial configuration $\mathbf{g}_0$ and random seed $\omega$, the equilibrium segregation index $S_\omega(T)$ is piecewise constant in $T$ with jumps only at values in $\mathcal{F}_8$. Write
\begin{equation}
S_\omega(T) = \sum_{\tau \in \mathcal{F}_8} \Delta S_\omega(\tau)\;\mathbf{1}[T \geq \tau]\,,
\label{eq:staircase_sum}
\end{equation}
where $\Delta S_\omega(\tau) = S_\omega(\tau^+) - S_\omega(\tau^-)$ is the jump at threshold $\tau$ (which depends on $\mathbf{g}_0$ and $\omega$).

With per-agent noise $\varepsilon_i \sim \mathcal{N}(0, \sigma^2)$, agent $i$'s effective threshold becomes $T_i = T + \varepsilon_i$. The indicator $\mathbf{1}[s_i < T_i]$ for agent $i$ being unsatisfied at nominal tolerance $T$ now depends on $\varepsilon_i$. Consider a threshold $\tau \in \mathcal{F}_8$. An agent whose noiseless satisfaction equals $\tau$ switches from satisfied to unsatisfied when $T + \varepsilon_i > \tau$, i.e., when $\varepsilon_i > \tau - T$. The probability of this event is
\begin{equation}
\mathbb{P}(\varepsilon_i > \tau - T) = \mathbb{P}\!\left(\frac{\varepsilon_i}{\sigma} > \frac{\tau - T}{\sigma}\right) = \Phi\!\left(\frac{T - \tau}{\sigma}\right).
\label{eq:switch_prob}
\end{equation}

For a given $(\mathbf{g}_0, \omega)$, the noisy segregation index is obtained by replacing each sharp jump $\mathbf{1}[T \geq \tau]$ with the smoothed version $\Phi((T - \tau)/\sigma)$:
\begin{equation}
\mathbb{E}_\varepsilon[S(T) \mid \mathbf{g}_0, \omega] = \sum_{\tau \in \mathcal{F}_8} \Delta S_\omega(\tau)\;\Phi\!\left(\frac{T - \tau}{\sigma}\right).
\label{eq:noise_avg}
\end{equation}

Taking the expectation over random initial configurations $\mathbf{g}_0$ and seeds $\omega$:
\begin{equation}
\langle S(T) \rangle = \sum_{\tau \in \mathcal{F}_8} \mathbb{E}[\Delta S_\omega(\tau)]\;\Phi\!\left(\frac{T - \tau}{\sigma}\right) + \text{Cov terms}.
\label{eq:full_avg}
\end{equation}

Define $w_\tau = \mathbb{E}[\Delta S_\omega(\tau)]$. The covariance terms arise because $\Delta S_\omega(\tau)$ and the noise realization are not independent across agents sharing neighbors. To bound them, write the segregation index as a sample mean $S = N^{-1}\sum_{i=1}^N h_i$, where $h_i$ depends on agent $i$'s noise $\varepsilon_i$ and the noises of its $\leq 8$ neighbors. Since the Moore neighborhood has diameter 1, agents at $\ell^\infty$-distance $\geq 3$ have independent $h_i$ values. The lattice can therefore be partitioned into $O(1)$ sublattices (at most $5^2 = 25$), each consisting of mutually independent agents. By the standard CLT for $m$-dependent random fields \citep{bolthausen1982}, the sum over each sublattice satisfies a CLT with rate $O(N^{-1/2})$, and the total covariance contribution is $O(N^{-1/2})$.

The ensemble averaging over different initial configurations $\mathbf{g}_0$ introduces additional fluctuations in $\Delta S_\omega(\tau)$. These act as a second source of broadening: the effective jump at threshold $\tau$ is not sharp but spread over a range $\pm\sigma_{\mathrm{ens}}$ reflecting the variability of $\mathbf{g}_0$. Since both broadenings are approximately Gaussian (the first by construction, the second by CLT over lattice sites), they combine in quadrature: $\sigma_{\mathrm{eff}}^2 = \sigma^2 + \sigma_{\mathrm{ens}}^2$, giving~\eqref{eq:ensemble}. \qed
\end{proof}

We fit~\eqref{eq:ensemble} to the $L = 320$ data using the 7 dominant $\mathcal{F}_8$ thresholds
\[
\tau \in \left\{\tfrac{1}{8},\; \tfrac{1}{4},\; \tfrac{2}{7},\; \tfrac{1}{3},\; \tfrac{3}{8},\; \tfrac{1}{2},\; \tfrac{5}{8}\right\}
\]
with free weights $\{w_\tau\}$ and a single shared $\sigma_{\mathrm{eff}}$. Non-negative least squares gives $\sigma_{\mathrm{eff}} = 0.032$ and $R^2 = 0.999$. The dominant weights are $w_{1/4} = 0.27$, $w_{1/2} = 0.22$, $w_{3/8} = 0.19$, $w_{1/8} = 0.12$. For comparison, a single error function achieves $R^2 = 0.265$ and a two-component version $R^2 = 0.997$.

A natural prediction for the weights is $w_\tau \propto \Delta\varphi(\tau)/(1 - R(\tau))$. This gets the ranking of the top two thresholds right but underestimates $w_{1/4}$ (predicted 0.037 vs.\ fitted 0.27). The reason is that $w_\tau$ measures the jump in \emph{segregation}, not in unsatisfied fraction: near $T = 1/4$ the system is close to well-mixed and each relocation produces a larger marginal increase in clustering. Deriving $\{w_\tau\}$ from the cascade dynamics would require tracking spatial correlations built up by successive relocations, which remains open.

\section{Beta Distribution for Heterogeneous Tolerance}
\label{app:beta}

We parameterize the agent-level tolerance distribution as $T_i \sim \mathrm{Beta}(\kappa/2, \kappa/2)$, giving density
\begin{equation}
f(t; \kappa) = \frac{t^{\kappa/2-1}(1-t)^{\kappa/2-1}}{B(\kappa/2, \kappa/2)}\,.
\label{eq:beta}
\end{equation}

We now derive the mean and variance. Write $\alpha = \beta = \kappa/2$. For a general $\mathrm{Beta}(\alpha, \beta)$ random variable, the $n$-th moment is
\begin{equation}
\mathbb{E}[T^n] = \frac{B(\alpha+n, \beta)}{B(\alpha, \beta)} = \prod_{r=0}^{n-1} \frac{\alpha + r}{\alpha + \beta + r}\,.
\label{eq:moments}
\end{equation}

The mean is therefore
\begin{equation}
\mathbb{E}[T] = \frac{\alpha}{\alpha + \beta} = \frac{\kappa/2}{\kappa/2 + \kappa/2} = \frac{\kappa/2}{\kappa} = \frac{1}{2}\,.
\label{eq:mean}
\end{equation}

For the variance, compute the second moment:
\begin{equation}
\mathbb{E}[T^2] = \frac{\alpha(\alpha+1)}{(\alpha+\beta)(\alpha+\beta+1)} = \frac{(\kappa/2)(\kappa/2+1)}{\kappa(\kappa+1)} = \frac{\kappa(\kappa+2)}{4\kappa(\kappa+1)} = \frac{\kappa+2}{4(\kappa+1)}\,.
\label{eq:secondmoment}
\end{equation}

Then
\begin{equation}
\mathrm{Var}(T) = \mathbb{E}[T^2] - (\mathbb{E}[T])^2 = \frac{\kappa+2}{4(\kappa+1)} - \frac{1}{4} = \frac{\kappa+2 - (\kappa+1)}{4(\kappa+1)} = \frac{1}{4(\kappa+1)}\,.
\label{eq:variance}
\end{equation}

At $\kappa = 2$ this reduces to the uniform distribution on $[0,1]$ (with variance $1/12$); as $\kappa \to \infty$, the variance vanishes and the distribution concentrates at $1/2$.

The symmetric parameterization ($\alpha = \beta = \kappa/2$) ensures that the two agent types face statistically identical tolerance distributions, isolating the effect of dispersion from any asymmetry in preferences. The single parameter $\kappa$ then has a clean interpretation: it measures how homogeneous the population is in its tolerance, independently of the mean tolerance level.

\section{Heterogeneous Tolerance Analysis}
\label{app:heterogeneous}

We replace the uniform tolerance $T$ by agent-specific thresholds $T_i \sim \mathrm{Beta}(\kappa/2, \kappa/2)$, where $\kappa > 0$ is a concentration parameter. The mean tolerance is always $1/2$, but the variance $1/(4(\kappa+1))$ ranges from broad ($\kappa = 1$: nearly uniform on $[0,1]$) to narrow ($\kappa = 100$: tightly concentrated at $1/2$).

\begin{figure}[!htb]
\centering
\includegraphics[width=\textwidth]{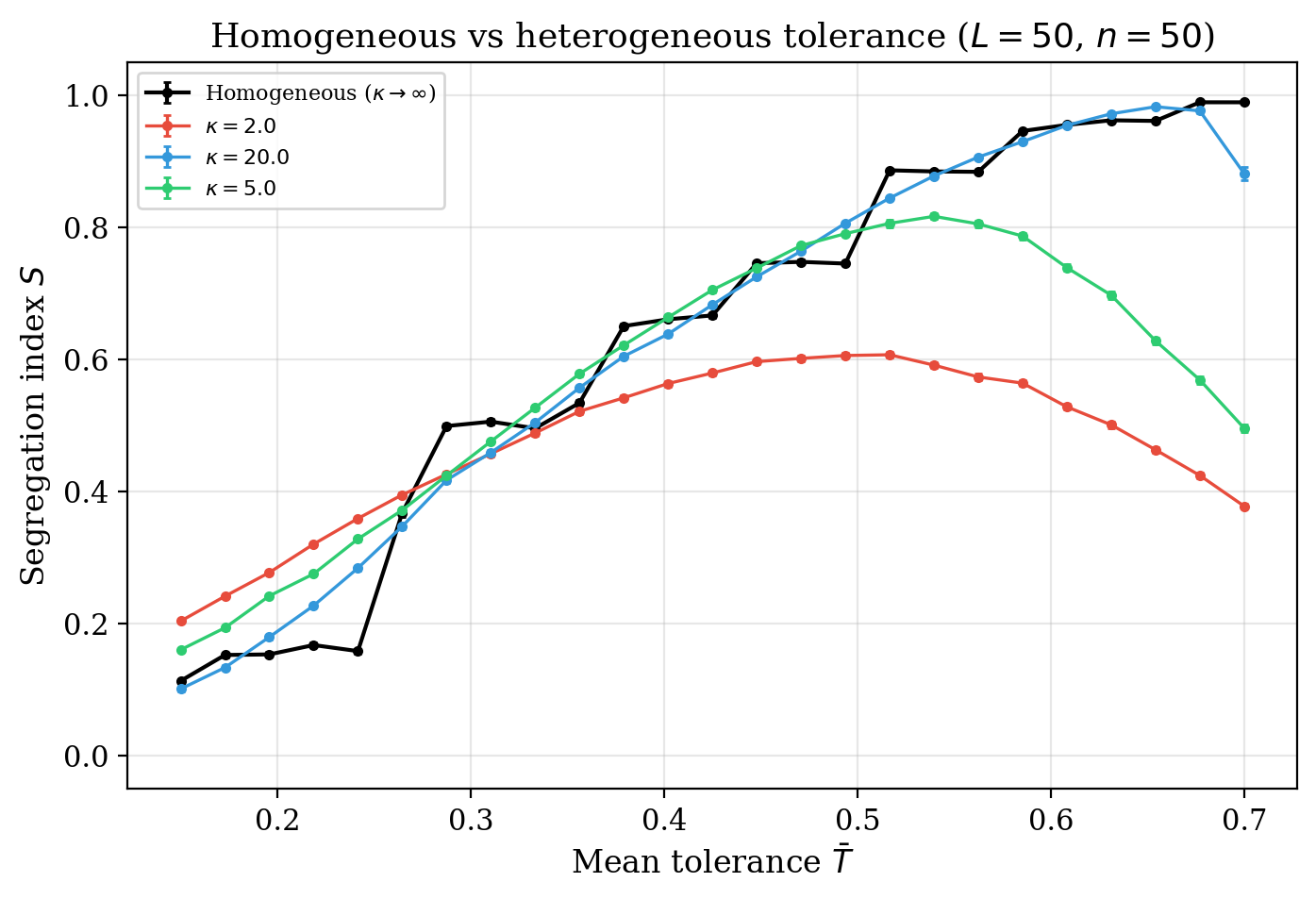}
\caption{Segregation curves for heterogeneous tolerance with $\kappa \in \{2, 5, 20\}$ and the homogeneous baseline. Lower $\kappa$ (wider spread) shifts the transition leftward and broadens it.}
\label{fig:heterogeneous}
\end{figure}

Figure~\ref{fig:heterogeneous} compares the segregation curves for $\kappa \in \{2, 5, 20\}$ against the homogeneous baseline. Two effects are apparent. First, dispersion smooths the transition: averaging over agents with different thresholds washes out the sharp collective switch. Second, and less obviously, the effective $T_c$ shifts \emph{downward} for small $\kappa$. The mechanism is that a dispersed population always contains a subpopulation of intolerant agents who begin segregating well before the mean threshold is reached, and their movement displaces other agents, triggering a cascade.

\FloatBarrier

\begin{figure}[!htb]
\centering
\includegraphics[width=\textwidth]{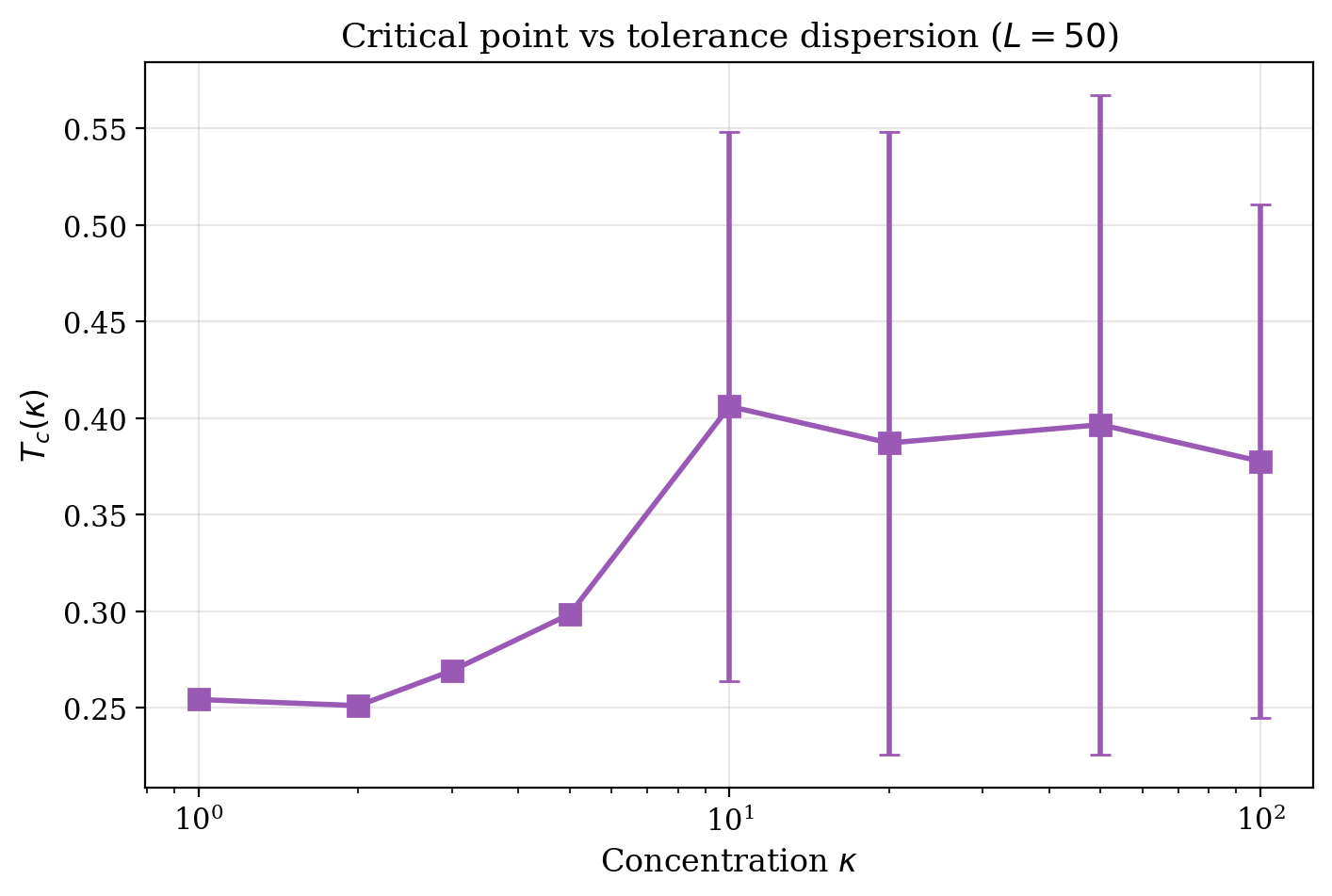}
\caption{Critical tolerance as a function of concentration $\kappa$. The transition sharpens and $T_c$ increases toward the homogeneous limit as $\kappa \to \infty$.}
\label{fig:heterogeneous_tc}
\end{figure}

Figure~\ref{fig:heterogeneous_tc} plots $T_c(\kappa)$ over the range $\kappa \in [1, 100]$. The critical tolerance increases from $\approx 0.25$ at $\kappa = 1$ to $\approx 0.40$ at $\kappa = 50$, then saturates. The convergence for large $\kappa$ is expected, since $\mathrm{Beta}(\kappa/2, \kappa/2) \to \delta(t - 1/2)$ as $\kappa \to \infty$. The broad error bars at intermediate $\kappa$ (10 to 50) reflect a genuinely wide transition region where the notion of a single $T_c$ becomes ambiguous.

\clearpage
\section{Numerical Parameters}
\label{app:params}

\captionof{table}{Parameters for the main finite-size scaling sweep.}
\label{tab:params}
\vspace{4pt}
\centering
\renewcommand{\arraystretch}{2.1}
{\Large
\begin{tabular}{ll}
\toprule
Parameter & Value \\
\midrule
Grid sizes $L$ & 20, 40, 80, 160, 320 \\
Density $\rho$ & 0.9 \\
Type-$A$ fraction $f_A$ & 0.5 \\
Tolerance range & $[0.1, 0.7]$, 50 equispaced points \\
Trials per $(L, T)$ pair & 50 \\
Maximum steps & 2000 \\
Convergence window & 20 steps \\
Convergence threshold & $10^{-3}$ \\
Per-agent tolerance noise $\sigma$ & 0.02 \\
\midrule
\multicolumn{2}{l}{\emph{Heterogeneous tolerance sweep}} \\
$\kappa$ values & 1, 2, 3, 5, 10, 20, 50, 100 \\
Tolerance range & $[0.1, 0.8]$, 30 points \\
Grid size & 80 \\
Trials per point & 50 \\
\bottomrule
\end{tabular}
}

\end{document}